\documentclass[pre,aps,superscriptaddress]{revtex4}

\usepackage[dvips]{graphics}
\usepackage{graphicx}
\usepackage{amsfonts}
\usepackage{amssymb}
\usepackage{amsmath}
\usepackage{subfigure}

\begin{document}

\title{Dark Solitons in Discrete Lattices: Saturable versus Cubic Nonlinearities}
\author{E.P. Fitrakis}
\affiliation{Department of Physics, University of Athens, Panepistimiopolis, Zografou,
Athens 15784, Greece }
\author{P. G.\ Kevrekidis}
\affiliation{Department of Mathematics and Statistics, University of
Massachusetts, Amherst MA 01003-4515}
\author{H.\ Susanto}
\affiliation{Department of Mathematics and Statistics, University of
Massachusetts, Amherst MA 01003-4515}
\author{D.J. Frantzeskakis}
\affiliation{Department of Physics, University of Athens, Panepistimiopolis, Zografou,
Athens 15784, Greece }
\
\begin{abstract}
In the present work, we study dark solitons in  
dynamical lattices with the saturable nonlinearity and compare them 
with
those in lattices 
with the cubic nonlinearity. This comparison has become especially 
relevant in light of recent experimental developments in the former
context. The stability properties of the fundamental waves, for both
on-site and inter-site modes, are examined analytically and
corroborated by numerical results. Furthermore, their dynamical 
evolution when they are found to be unstable is obtained through
appropriately crafted numerical experiments.
\end{abstract}

\maketitle

\section{Introduction.} 

In the past few years, the investigation of 
dynamical nonlinear lattices, where the ``evolution variable''
is continuum, while the ``spatial variables'' are discrete, 
has seen a tremendous growth. 
One of the main directions that has triggered both experimental and theoretical efforts in this context  
has been the development of optically induced lattices in photorefractive media, such as Strontium Barium Niobate
(SBN) \cite{efrem}, and its experimental realization \cite{moti1,moti2}. As a result, a remarkable set of 
nonlinear waves has been predicted and experimentally observed; these include
(but are not limited to) discrete dipole \cite{dip}, quadrupole \cite{quad}, necklace \cite{neck} and other
multi-pulse/multi-pole localized patterns \cite{multi}, impurity modes \cite{fedele}, discrete vortices 
\cite{vortex}, and solitons in radial lattices \cite{rings}.
A recent review of this direction can be found in \cite{moti3}.

Another direction, also relevant to nonlinear optics, which has been 
growing in parallel, concerns the intriguing interplay of 
nonlinearity and discrete diffraction emerging in fabricated AlGaAs 
waveguide arrays \cite{7}. A variety of phenomena, such as discrete diffraction, 
Peierls barriers, diffraction management \cite{7a} and the formation of gap solitons \cite{7b} among
others, have been experimentally traced in the latter setting. 
These phenomena, in turn, stimulated a tremendous increase in the number of the theoretical 
studies concerned with such effectively discrete media
(see, e.g., the reviews \cite{review_opt,general_review}). 

Additionally, in the past decade, there has been yet another context where the  
consideration of discrete lattices and nonlinear waves thereof 
are relevant; this is the soft-condensed matter physics 
of Bose-Einstein Condensates (BECs) \cite{books}. Droplets 
of such condensates may be trapped in 
a periodic optical potential commonly known as ``optical lattice'' (see, e.g., \cite{ibloch} and references therein). 
The physics of BECs confined in optical lattices has also experienced a tremendous growth over the past few years, 
leading to many important theoretical and experimental results; these 
include the prediction and manifestation of modulational instabilities
\cite{pgk}, the observation of gap solitons \cite{markus} and Landau-Zener
tunneling \cite{arimondo} among many other features; reviews of
the theoretical and experimental findings in this area have also 
recently appeared in \cite{konotop,markus2}.

Many of the above studies of coherent structures (especially, as regards 
one-dimensional ones) have been conducted in self-focusing media (i.e., media
with ``attractive'' nonlinearities), where bright solitons or multi-soliton
versions thereof emerge. Dark solitons in lattices with the 
so-called self-defocusing (or ``repulsive'', in the language of BECs) 
nonlinearity have, in some sense, been far less popular with only a few 
examples of relevant experimental studies. In particular, the first
observation (to the best of our knowledge) of dark solitons arose
in \cite{silb} in the anomalous diffraction regime of AlGaAs waveguide arrays, which feature 
the Kerr-type cubic nonlinearity. On the other hand, very recently, dark solitons were realized in defocusing lithium
niobate waveguide arrays, which exhibit a different type of nonlinearity,
namely a saturable, defocusing one due to the photovoltaic effect \cite{kip}.

The scope of the present work is to compare 
the above settings, namely that of the cubic nonlinearity
with that of the saturable one, regarding the properties of dark solitons.
In particular, our aim here is to examine both of these models and make 
inferences
on the properties of the dark solitons on the basis of existence 
and stability analysis. We aim to mathematically justify the stability
of the pertinent dark soliton modes (such as those centered on a site
and those centered between two lattice sites) and to numerically corroborate
our analytical findings. Furthermore, we intend to examine more quantitatively 
 statements, such as those of \cite{kip} about improved stability
of inter-site modes, and to examine whether phenomena similar to those
emerging for bright solitons (such as the stability alternation between
on-site and inter-site solutions \cite{kip2}) can occur for dark solitons. 
Our work also builds on earlier theoretical work on such structures which 
considered them principally in the cubic nonlinearity setting such as 
e.g., \cite{oksana,joh,hs}. We expand the considerations of these earlier works by considering
in more detail the stability problem and obtaining information 
about the key eigenvalues (and doing that systematically, starting from 
the anti-continuum limit, where the lattice sites are uncoupled), 
as well as applying these ideas to a different setting, such as that of the 
saturable nonlinearity and readily comparing the results of the two cases.

The paper is organized as follows. In the next section, we will present our analytical considerations both for the 
saturable and for the cubic case. Subsequently, we will focus on numerical results and examine how they compare with
the analytical predictions. In the fourth 
section, we will summarize our findings, 
present our conclusions, and indicate some interesting directions for future studies. Finally, the appendix contains an alternative, 
perturbation-theory based description of our analytical results that we 
will illustrate to 
be equivalent (for small couplings) to the analysis of the earlier sections
(and is included for reasons of completeness of the mathematical part of
our exposition).

\section{Models and Theoretical Setup}

\subsection{Cubic Nonlinearity}

We will start the discussion of our analytical results by
considering the somewhat simpler (at the level of mathematical
details) case of the lattice exhibiting the cubic nonlinearity. 
The relevant model is the discrete nonlinear Schr\"{o}dinger (NLS) equation of the following dimensionless form:
\begin{eqnarray}
i \dot{u}_n=-\epsilon \Delta_2 u_n + \beta_1 |u_n|^2 u_n,
\label{ceq1}
\end{eqnarray}
where $u_n$ denotes the complex electric field envelope 
in the case of waveguide arrays, or the wavefunction of the BEC droplets 
in the wells of a deep optical lattice. The parameter $\epsilon$ 
is the coupling constant that characterizes the tunneling
coefficient, while the parameter $\beta_1 > 0$ (which can be scaled out but
will be kept for the purposes of completeness/comparison
with the saturable model) sets the strength of the nonlinearity.
The overdot denotes differentiation with respect to the evolution
variable $t$ (which is the propagation distance in optics, and time in BECs), 
and, finally, $\Delta_2 u_n = (u_{n+1}+u_{n-1}-2 u_n)$ 
stands for the discrete Laplacian.
Notice that we will be, in particular, concerned with
steady state solutions of the form 
\begin{eqnarray}
u_n=v_n \exp(-i \Lambda_1 t),
\label{ceq2}
\end{eqnarray}
where $\Lambda_1$ denotes the propagation constant (or the chemical potential in the BEC context). 
This leads to the steady state equation:
\begin{eqnarray}
\beta_1 |v_n|^2 v_n - \Lambda_1 v_n - \epsilon \Delta_2 v_n = 0. 
\label{ceq3}
\end{eqnarray}
We will also be concerned with the linear stability of the ensuing
solutions which can be studied using the ansatz
\begin{eqnarray}
u_n=\exp(-i \Lambda_1 t) \left[v_n + \delta \left(\exp(\lambda t) p_n 
+ \exp(\lambda^{\star} t) q_n \right) \right],
\label{ceq4a}
\end{eqnarray}
where $\delta$ is a formal linearization parameter and the resulting
linear problem (to O$(\delta)$) 
for the eigenvalue $\lambda$ and the eigenvector $(p_n,q_n^{\star})^T$  
determines the stability of the configuration as follows: 
a configuration will be (neutrally) stable for this Hamiltonian system if 
$\forall \lambda$, the real part $\lambda_r$ of the eigenvalue
($\lambda=\lambda_r + i \lambda_i$) is such that $\lambda_r=0$.
This is due to the fact that the Hamiltonian nature of the system
enforces that if $\lambda$ is an eigenvalue, so are $\lambda^{\star}$, 
$-\lambda$ and $-\lambda^{\star}$. Note that in the above expressions $^{\star}$ 
denotes complex conjugate, while $^T$ denotes transpose. 
The linear stability problem can be written in compact notation in the form:
\begin{eqnarray}
i \lambda \left( \begin{array}{c} p_n \\ q_n^{\star}
\end{array} \right) = \left( \begin{array}{cc} 
2 \beta_1 |v_n|^2 -\Lambda_1 - \epsilon \Delta_2 & \beta_1 v_n^2 \\ 
-\beta_1 (v_n^2)^{\star} & \Lambda_1 -2 \beta_1 |v_n|^2 + \epsilon \Delta_2
\end{array} \right) \left( \begin{array}{c} p_n \\ q_n^{\star}
\end{array} \right),
\label{ceq4b}
\end{eqnarray}

\subsubsection{Anti-Continuum Limit}

The anti-continuum (AC) limit of the equation is characterized by $\epsilon=0$. 
Then, the steady states are straightforwardly characterized by:
\begin{eqnarray}
v_n = \left\{\sqrt{\frac{\Lambda_1}{\beta_1}},0 \right \} \exp(i \theta_n),
\label{ceq4}
\end{eqnarray}
where $\theta_n$ is an arbitrary phase. We will restrain
ourselves here (without loss of generality, see e.g., \cite{dep})
to real solutions with $\theta_n \in \{0,\pi\}$.

We will be concerned with two types of dark soliton solutions,
the well-known on-site and inter-site one (see e.g. \cite{oksana,joh}).
The former one, at the AC-limit, is of the form:
\begin{eqnarray}
v_{n \leq -1} &=& \sqrt{\frac{\Lambda_1}{\beta_1}},
\label{ceq5}
\\
v_{n = 0} &=& 0,
\label{ceq6}
\\
v_{n \geq 1} &=& -\sqrt{\frac{\Lambda_1}{\beta_1}}, 
\label{ceq7}
\end{eqnarray}
while the latter type of dark soliton reads:
\begin{eqnarray}
v_{n \leq 0} &=& \sqrt{\frac{\Lambda_1}{\beta_1}}
\label{ceq5a}
\\
v_{n \geq 1} &=& -\sqrt{\frac{\Lambda_1}{\beta_1}}
\label{ceq7a}
\end{eqnarray}

One can subsequently examine the linear stability of these
prototypical configurations, as a starting point 
for the finite $\epsilon$ case. In the AC limit, the linear 
stability simplifies greatly due to the fact that the sites 
become uncoupled. One can then easily see that for all 
non-zero sites, the relevant stability matrix of the AC-limit 
is identical and has the form:
\begin{eqnarray}
\Lambda_1 \left( \begin{array}{cc} 
 1 &  1 \\ 
-1 & -1 \end{array} \right)
\label{ceq8}
\end{eqnarray}
This yields a pair of zero eigenvalues for each of these non-zero
sites. Hence, in an inter-site configuration at the AC-limit, 
the linearization would only result in such zero eigenvalues.

The only difference of an on-site configuration lies in the
existence of the central $v_0=0$ site. The latter produces
a $2 \times 2$ stability matrix of the form
\begin{eqnarray}
\Lambda_1 \left( \begin{array}{cc} 
 -1 &  0 \\ 
  0 &  1  \end{array} \right),
\label{ceq8a}
\end{eqnarray}
and, therefore, an eigenvalue pair $\lambda = \pm i \Lambda_1$.

\subsubsection{Finite Coupling Case}

Let us first consider the solution profile. It is clear that
for finite $\epsilon$ the solutions will be deformed from their
AC-limit profile of Eqs. (\ref{ceq5})-(\ref{ceq7a}). To address
this deformation, we can expand the steady state solution into a power series:
\begin{eqnarray}
v_n=v_n^{(0)} + \epsilon v_n^{(1)} + {\rm O}(\epsilon^2).
\label{ceq9}
\end{eqnarray}
The leading order correction can be straightforwardly computed
by the substitution of the expansion into Eq. (\ref{ceq3}), as
\begin{eqnarray}
v_n^{(1)}=\frac{\Delta_2 v_n^{(0)}}{2 \Lambda_1}
\label{ceq10}
\end{eqnarray}
for all excited sites. For the zeroth site of the on-site
configuration, the symmetry of the profile yields a zero 
correction (to all relevant orders). It is easy to see that
the correction of Eq. (\ref{ceq10}) only contributes to leading
order to the sites with $n \in \{1,-1\}$ for the on-site 
and to those with $n \in \{0,1\}$ for the inter-site configuration.
These corrections amount to:
\begin{eqnarray}
v_1^{(1)} &=&\frac{1}{2 \sqrt{\Lambda_1 \beta_1}}
\label{ceq11}
\\
v_{-1}^{(1)} &=& -v_1^{(1)}
\label{ceq12}
\end{eqnarray}
for the on-site and to:
\begin{eqnarray}
v_1^{(1)} &=&\frac{1}{\sqrt{\Lambda_1 \beta_1}}
\label{ceq11a}
\\
v_{0}^{(1)} &=& -v_1^{(1)}
\label{ceq12a}
\end{eqnarray}
for the inter-site.

The next step is to consider the stability problem. Clearly, the
latter will have eigenvalue contributions from two sources.
One will be the continuous spectrum that will emerge from the
background (that corresponded to zero eigenvalues at the AC limit)
and the other will be from the point spectrum in the vicinity
of the center of the dark soliton configuration. We commence from
the easier calculation of the former type. The continuous spectrum
corresponds to plane wave eigenfunctions of the type 
$\{p_n,q_n\} \sim \exp(i k n)$. These, in turn, result into
a matrix eigenvalue problem, where the matrix is of the form:
\begin{eqnarray}
 \left( \begin{array}{cc} 
 s_1 &  \Lambda_1 \\ 
-\Lambda_1 & -s_1 \end{array} \right),
\label{ceq13}
\end{eqnarray}
where $s_1=\Lambda_1+4 \epsilon \sin^2(k/2)$. This leads to
a continuous eigenvalue spectrum described by the dispersion relation 
\begin{eqnarray}
\omega \equiv i \lambda = \pm \sqrt{s_1^2-\Lambda_1^2},
\label{ceq14}
\end{eqnarray}
which is associated with the eigenvalue band 
$i \lambda \in [-\sqrt{16 \epsilon^2+8 \epsilon \Lambda_1},\sqrt{16 \epsilon^2
+8 \epsilon \Lambda_1}]$.

We now turn to the point spectrum eigenvalues, stemming from the
central part of the excitation. For the inter-site configuration, in order
to address this issue, we consider to a leading order approximation
only the two sites ($n=0$ and $n=1$) participating in the dark
soliton (as they are the only ones modified to leading order in
perturbation theory). Using the perturbative expansion of Eqs. 
(\ref{ceq5a})-(\ref{ceq7a}) within the relevant part of the eigenvalue
problem, we obtain the $4 \times 4$ matrix
\begin{eqnarray}
 \left( \begin{array}{cccc} 
 \Lambda_1-2 \epsilon &  -\epsilon & \Lambda_1-2 \epsilon &  0 \\
 -\epsilon & \Lambda_1-2 \epsilon & 0 & \Lambda_1 -2 \epsilon \\
 -\Lambda_1+2 \epsilon & 0 & -\Lambda_1+2 \epsilon & \epsilon  \\
 0 & -\Lambda_1+2 \epsilon & \epsilon & -\Lambda_1 + 2 \epsilon 
\end{array} \right),
\label{ceq15}
\end{eqnarray}
which, importantly, leads to a pair of real eigenvalues 
\begin{eqnarray}
\lambda= \pm \sqrt{2 \epsilon \Lambda_1- 5 \epsilon^2},
\label{ceq16}
\end{eqnarray}
which render the configuration immediately unstable, as soon as 
$\epsilon$ becomes non-zero. This prediction will be compared with 
the numerical results in the following section.

Finally, for the onsite configuration, one can use a similar 
argument entailing the three central sites of the solitary structure 
and constructing a $6 \times 6$ matrix whose eigenvalues can, in 
principle, be computed. However, the resulting expressions are 
too cumbersome to be useful, hence, here we offer two alternative 
arguments that, as we will show, describe very accurately the
behavior of the relevant point spectrum eigenvalue originally
at $\lambda = \pm i \Lambda_1$, associated with the on-site soliton. 

The first approach that we offer is a rigorous one and
is based on the so-called Gerschgorin's theorem \cite{atkinson}.
Let us consider matrices $A=[a_{lj}]$ of order $N$ and define the radii
$r_l=\sum_{j=1, j\neq l}^N |a_{lj}|$ and denote the circles in the complex
spectral plane $Z_l=\{z \in C: |z-a_{ll}|<r_l\}$. Then, Gerschgorin's
theorem states that the eigenvalues of the matrix belong to these
circles and, in fact, its refined version states that if $m$ of these
circles form a connected set, $S$, disjoint from the remaining
$N-m$ circles, then exactly $m$ eigenvalues are contained in $S$.
Our problem is an excellent testbed for the application of
Gerschgorin's theorem because the sole eigenvalue discussed above
is at $\pm i \Lambda_1$ in the AC-limit (of zero radius for the
Gerschgorin circles), while all others are located at the origin.
Hence, for small $\epsilon \ll \Lambda_1$, the (single) relevant point
spectrum eigenvalue remains in the corresponding Gerschgorin 
circle which can be easily computed. In fact, considering that
only the diagonal and super- and sub-diagonal elements are present
for a site with $v_0=0$, the Gerschgorin estimate for the
relevant eigenvalue emerges immediately as:
\begin{eqnarray}
|i \lambda\pm \left(\Lambda_1-2 \epsilon \right)| \leq 2 \epsilon
\label{ceq17}
\end{eqnarray}
which, in turn, necessitates that the relevant eigenvalue lies
between $\Lambda_1-4 \epsilon \leq i \lambda \leq \Lambda_1$.
This is a rigorous result based on the above theorem, which provides
a definitive (linear) bound on the growth of the relevant eigenvalue.

On the other hand, there is also an even more successful approach
(which, however, is to a certain degree an approximation), according
to which, one considers the eigenvalue equations for the eigenvectors
$p_n$ and $q_n^{\star}$. Considering the anti-symmetry of the central
site one may use the approximation of $p_{n+1}+p_{n-1} \approx 0$
(in lines similar to \cite{joh,hs}, which, however, postulated 
the stronger condition $p_{n+1} \approx p_{n-1} \approx 0$). In the latter 
setting, the eigenvalue equations (\ref{ceq4b}) decouple and provide
a relevant estimate for this eigenvalue as
\begin{eqnarray}
i \lambda = \pm \left(\Lambda_1 -2 \epsilon \right).
\label{ceq18}
\end{eqnarray}
This prediction will also be compared with our numerical results in
the next section. It is also worth noting here that this eigenvalue,
moving towards the spectral plane origin, will collide with the
band edge of the continuous spectrum when the predictions of
Eq. (\ref{ceq18}) and of the band edge of the continuous spectrum
coincide, which occurs for
\begin{eqnarray}
\epsilon_{cr} =\frac{2 \sqrt{3}-3}{6} \Lambda_1 \equiv 0.07735 \Lambda_1.
\label{ceq19}
\end{eqnarray}
This collision, per the opposite Krein signature \cite{joh2,krein} of
the relevant eigenvalues, will lead to a Hamiltonian Hopf bifurcation
\cite{HH} and a quartet of eigenvalues, similarly to what was found
in \cite{joh} (for relevant details, see below).

\subsection{Saturable Nonlinearity}

The second model that we will consider is a discrete dynamical lattice exhibiting a saturable defocusing 
nonlinearity \cite{kip2,hadz1}, which has recently been used in a variety
of studies and pertains to the 
dark soliton experiments reported in \cite{kip}. In this case, the relevant discrete NLS equation is 
of the form:
\begin{eqnarray}
i \dot{u}_n=-\epsilon \Delta_2 u_n - \frac{\beta_2}{1+|u_n|^2} u_n, 
\label{ceq20}
\end{eqnarray}
where the nonlinearity parameter $\beta_2 > 0$. 
Similarly to the previous case, we use the standing wave ansatz of the form $u_n=\exp(i \Lambda_2 t) v_n$,
to obtain the steady state equation
\begin{eqnarray}
\Lambda_2 v_n  -  \frac{\beta_2}{1+|v_n|^2} v_n - \epsilon \Delta_2 v_n = 0.
\label{ceq21}
\end{eqnarray}
Using a similar ansatz to that of Eq. (\ref{ceq4a}), we then obtain the following stability equations: 
\begin{eqnarray}
i \lambda \left( \begin{array}{c} p_n \\ q_n^{\star}
\end{array} \right) = \left( \begin{array}{cc} 
\Lambda_2 - \frac{\beta_2}{1+|v_n|^2}+\frac{\beta_2 |v_n|^2}{(1+|v_n|^2)^2} - \epsilon \Delta_2 & \frac{\beta_2 v_n^2}{(1+|v_n|^2)^2} \\ 
-\frac{\beta_2 (v_n^2)^{\star}}{(1+|v_n|^2)^2} & -\Lambda_2 + \frac{\beta_2}{1+|v_n|^2} -\frac{\beta_2 |v_n|^2}{(1+|v_n|^2)^2} + \epsilon \Delta_2
\end{array} \right) \left( \begin{array}{c} p_n \\ q_n^{\star}
\end{array} \right).
\label{ceq21b}
\end{eqnarray}

\subsubsection{Anti-Continuum Limit}

As before, we commence our analysis of the model at the AC-limit of
$\epsilon=0$, where the solutions can be found explicitly in the form
\begin{eqnarray}
v_n=\left\{\sqrt{\frac{\beta_2}{\Lambda_2}-1},0 \right \} \exp(i \theta_n),
\label{ceq22}
\end{eqnarray}
where we will assume, without loss of generality, that $\Lambda_2 
\leq \beta_2$. As in the case of the cubic nonlinearity, the on-site dark soliton of the
AC-limit has the form
\begin{eqnarray}
v_{n \leq -1} &=& \sqrt{\frac{\beta_2}{\Lambda_2}-1}
\label{ceq5b}
\\
v_{n = 0} &=& 0
\label{ceq6b}
\\
v_{n \geq 1} &=& -\sqrt{\frac{\beta_2}{\Lambda_2}-1},
\label{ceq7b}
\end{eqnarray}
while the inter-site one reads
\begin{eqnarray}
v_{n \leq 0} &=& \sqrt{\frac{\beta_2}{\Lambda_2}-1}
\label{ceq5c}
\\
v_{n \geq 1} &=& -\sqrt{\frac{\beta_2}{\Lambda_2}-1}
\label{ceq7c}
\end{eqnarray}

Now, we turn to the stability problem in the AC-limit. Once
again, there are two types of blocks in the stability matrix,
one associated with sites belonging to the background,
and another related to the central zero site of the on-site 
configuration. The former $2 \times 2$ block is of the form
\begin{eqnarray}
\frac{\Lambda_2^2}{\beta_2} \left(\frac{\beta_2}{\Lambda_2}-1\right) 
\left( \begin{array}{cc} 
 1 &  1 \\ 
-1 & -1 \end{array} \right),
\label{ceq8b}
\end{eqnarray}
leading once again to zero eigenvalues for the excited sites.
On the other hand, for the site with $v_0=0$, the related
block is:
\begin{eqnarray}
\left( \Lambda_2-\beta_2\right ) \left( \begin{array}{cc} 
 1 &  0 \\ 
  0 &  -1  \end{array} \right),
\label{ceq8c}
\end{eqnarray}
leading to an eigenvalue pair $\lambda=\pm i (\Lambda_2-\beta_2)$,
which is again isolated from the rest of the (zero) eigenvalues.

\subsubsection{Finite Coupling Case}

For the finite coupling case, we once again expand the solution
in a power series in $\epsilon$ as in Eq. (\ref{ceq9}) and obtain
the leading order correction corresponding to the excited sites as:
\begin{eqnarray}
v_n^{(1)}=\frac{\Delta_2 v_n^{(0)}}{\frac{2 \beta_2 (v_n^{(0)})^2}{1+(v_n^{(0)})^2}}
\label{ceq8d}
\end{eqnarray}
This results into the leading order corrections for the on-site mode:
\begin{eqnarray}
v_1^{(1)} &=&\frac{\beta_2}{2 \Lambda_2^2 \sqrt{\frac{\beta_2}{\Lambda_2}-1}}
\label{ceq8e}
\\
v_{-1}^{(1)} &=& -v_1^{(1)},
\label{ceq8f}
\end{eqnarray}
while for the inter-site mode:
\begin{eqnarray}
v_1^{(1)} &=&\frac{\beta_2}{\Lambda_2^2 \sqrt{\frac{\beta_2}{\Lambda_2}-1}}
\label{ceq8h}
\\
v_{-1}^{(1)} &=& -v_1^{(1)}.
\label{ceq8i}
\end{eqnarray}

Having examined the leading order perturbation to the solution, we now
focus on the stability for the finite coupling case. The continuous spectrum
can be found, as before, by assigning $\{p_n,q_n\} \sim \exp(i k n)$;
using the notation $s_2=(\Lambda_2^2/\beta_2) (\beta_2/\Lambda_2-1)$,
as well as $\tilde{s}_2=s_2+4 \epsilon \sin^2(k/2)$, the relevant matrix becomes 
\begin{eqnarray}
 \left( \begin{array}{cc} 
 \tilde{s}_2 &  s_2 \\ 
-s_2 & -\tilde{s}_2 \end{array} \right),
\label{ceq8j}
\end{eqnarray}
which yields the continuous spectrum 
\begin{eqnarray}
\omega \equiv i \lambda = \pm \sqrt{\tilde{s}_2^2-s_2^2},
\label{ceq8k}
\end{eqnarray}
and the phonon band $i \lambda \in [-\sqrt{16 \epsilon^2+8 \epsilon s_2},
\sqrt{16 \epsilon^2+8 \epsilon s_2}]$.

Finally, we consider the point spectrum associated with the inter-site
and the on-site configurations. For the former, once again we 
consider the $4 \times 4$ matrix block associated with the two
central sites. However, given the additional complexity of the 
saturable case, in order to obtain an analytically tractable 
leading order expression, we make a further simplification here.
Namely, in all the elements of the matrix that have O$(1)$ terms
with O$(\epsilon)$ corrections, we only consider the O$(1)$ terms.
This leads to the following matrix
\begin{eqnarray}
 \left( \begin{array}{cccc} 
 s_2 &  -\epsilon & s_2 &  0 \\
 -\epsilon & s_2 & 0 & s_2 \\
 -s_2 & 0 & -s_2 & \epsilon  \\
 0 & -s_2 & \epsilon & -s_2 
\end{array} \right),
\label{ceq8l}
\end{eqnarray}
which, importantly, leads to a pair of real eigenvalues
\begin{eqnarray}
\lambda= \pm \sqrt{2 \epsilon s_2- \epsilon^2},
\label{ceq8m}
\end{eqnarray}
that will be compared with our numerical findings.

On the other hand, for the onsite case, similar considerations to
those of the cubic case simplify the results considerably (despite
the additional complications of the saturable case linearization).
In particular, we can still apply the Gerschgorin argument to obtain
a rigorous bound on the eigenvalue associated with the zero site as
\begin{eqnarray}
|i \lambda\pm \left(\Lambda_2-\beta_2-2 \epsilon \right)| \leq 2 \epsilon.
\label{ceq8n}
\end{eqnarray}
This results in the inequality $\beta_2-\Lambda_2-4 \epsilon \leq i 
\lambda \leq \beta_2-\Lambda_2$.

However, again a very accurate description (as will be evinced by
our numerical computations below) can be provided by the
anti-symmetric approximation of $p_{n+1}+p_{n-1} \approx 0$, which
leads to the eigenvalue pair:
\begin{eqnarray}
i \lambda = \pm \left( \beta_2 - \Lambda_2 - 2 \epsilon \right).
\label{ceq8o}
\end{eqnarray}
Lastly, the collision of this pair with the continuous spectrum
band derived above is what will produce the instability, hence equating
the above expression with that of Eq. (\ref{ceq8k}), we obtain the
critical coupling of 
\begin{eqnarray}
\epsilon_{cr} = \frac{1}{6} 
\left[2 \sqrt{s_2^2+(\beta_2-\Lambda_2)^2 + (\beta_2-\Lambda_2) s_2}
-2 s_2 - (\beta_2-\Lambda_2) \right].
\label{ceq8p}
\end{eqnarray}

Before we move to our numerical results and their comparison
with our analytical findings let us point out here that the 
main stability results obtained above will also be confirmed
in the Appendix using a different mathematical technique.
In particular, for completeness in the Appendix we consider
a perturbative formulation of the stability problem and identify
the key eigenvalues through their formal series expansion
in powers of $\epsilon$ (similarly to what was done above for
the corrections to the solution itself). In this alternative
way, we will show how to obtain the leading order analogs of
Eqs. (\ref{ceq16}) [for cubic, inter-site], (\ref{ceq18}) 
[for cubic on-site], (\ref{ceq8m}) [for saturable, inter-site]
and (\ref{ceq8o}) [for saturable, on-site].

\section{Numerical Results and Comparison}

In our numerical results we will consider the solutions to the two models
as a function of the coupling constant $\epsilon$, for the following
selection of parameters $(\beta_1,\Lambda_1)=(1,1)$ for the cubic
case and $(\beta_2,\Lambda_2)=(1,0.5)$ for the saturable one.
Recall that $\beta_1$ and $\beta_2$, the parameters characterizing
the strength of the nonlinear terms, can always be scaled out 
with a rescaling of $\epsilon$ and a re-parametrization
of time. This indicates that they can be chosen to be equal to unity; 
this way, the two models also possess the same type of nonlinearity for small 
amplitude solutions (as can be readily observed by Taylor expanding
the saturable model for $|u|^2 \ll 1$). In the cubic nonlinear model, $\Lambda_1$
can also be absorbed in a rescaling of $\epsilon$ and of the solution
amplitude, hence we chose a typical value pertaining to a unity 
background (but our calculations above and some discussion below 
illustrate how the results depend on $\Lambda_1$).
Given the above three choices, $\Lambda_2$ is then chosen in a way 
such that the two models have the same background intensity. 
Generally, the latter requirement imposes the condition
\begin{eqnarray}
\frac{\Lambda_1}{\beta_1}=\frac{\beta_2}{\Lambda_2}-1,
\label{ceq23}
\end{eqnarray}
to which we return below.

\begin{figure}[tbp]
\begin{center}
\hskip-0.15cm
\begin{tabular}{cc}
\includegraphics[height=6cm,width=6cm]{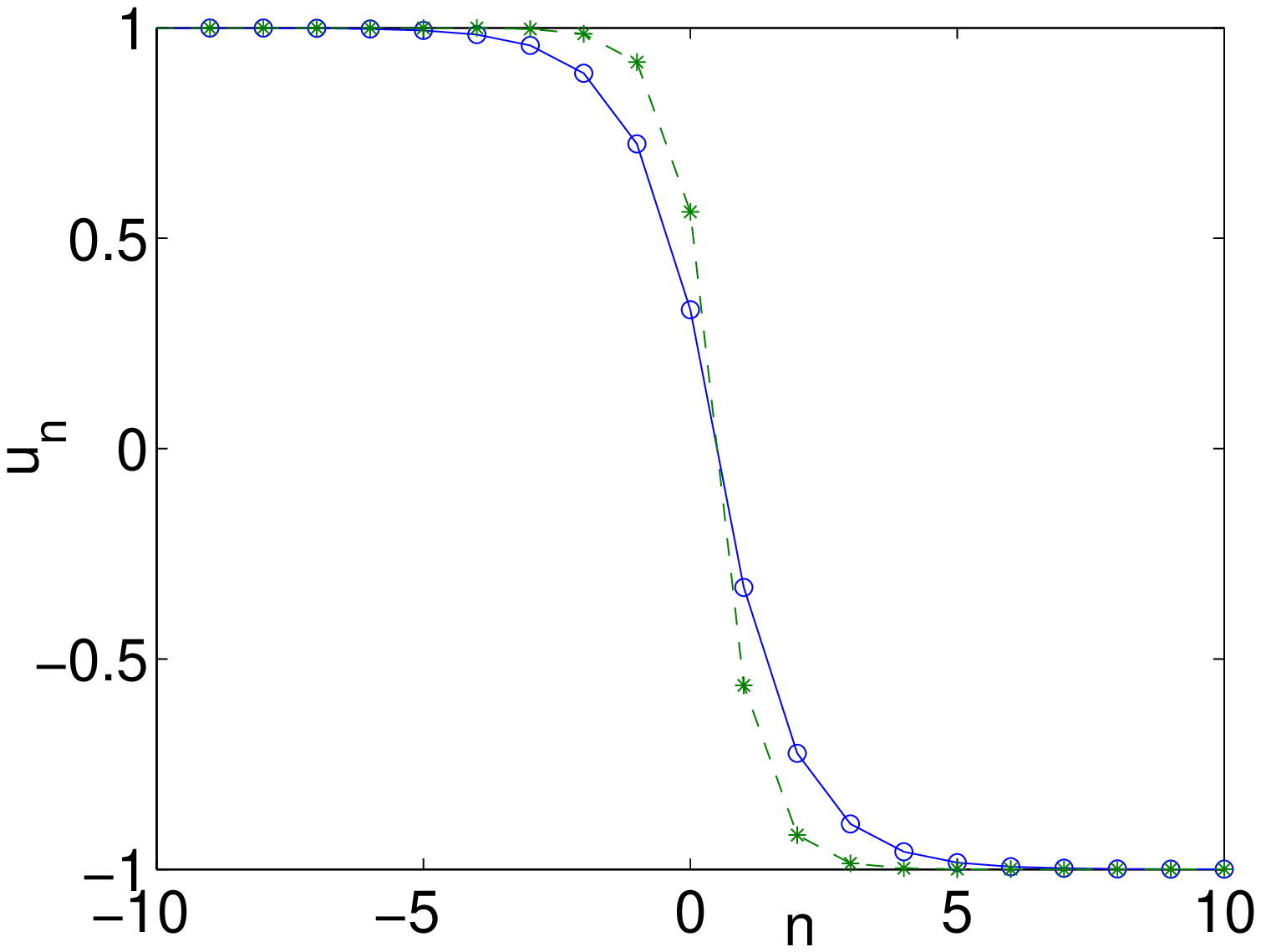}
\includegraphics[height=6cm,width=6cm]{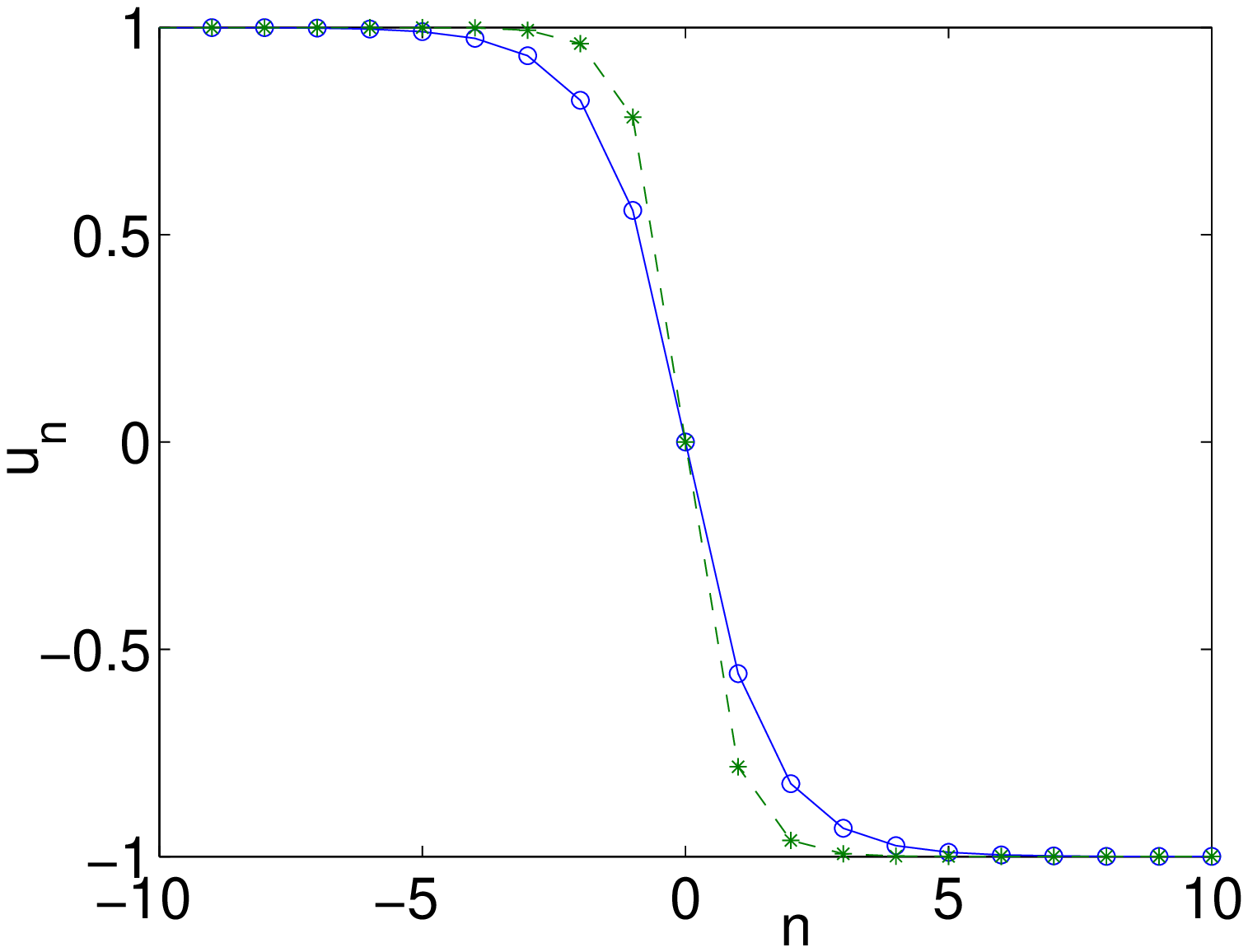} &  \\
\end{tabular}\end{center}
\par
\vskip-0.7cm 
\caption{(Color online) Profiles of the dark soliton solutions of the two
models for unity background and $\epsilon=0.5$ as a function of the
spatial grid variable $n$. The left panel shows the inter-site solutions,
while the right one shows the on-site solutions. The solutions to the cubic
model are marked by stars, connected (as a guide to the eye) by a dashed
line, while those to the saturable model are shown by circles, connected
by a solid line.}
\label{fig1}
\end{figure}

Firstly, in Fig. \ref{fig1} we show a typical comparison of the numerically
obtained inter-site, as well as on-site exact (up to a prescribed
numerical tolerance) solutions for a typical value of $\epsilon$
(namely, $\epsilon=0.5$). These results have been
obtained by means of the Newton-Raphson root-finding algorithm, as fixed points of 
a numerical iteration scheme on the lattice grid.
The stars that are connected (as a guide to the eye) by dashed lines pertain to the cubic model,
as opposed to the circles (connected by a solid line) for the
saturable model. It can easily be observed that the former
kinks are narrower and steeper, which can be attributed to
the ``stronger'' nature of the cubic nonlinearity (in comparison
with the weaker saturable one, which returns to a linear behavior
for large amplitudes). 

\begin{figure}[tbp]
\begin{center}
\hskip-0.15cm
\begin{tabular}{cc}
\includegraphics[height=6cm,width=6cm]{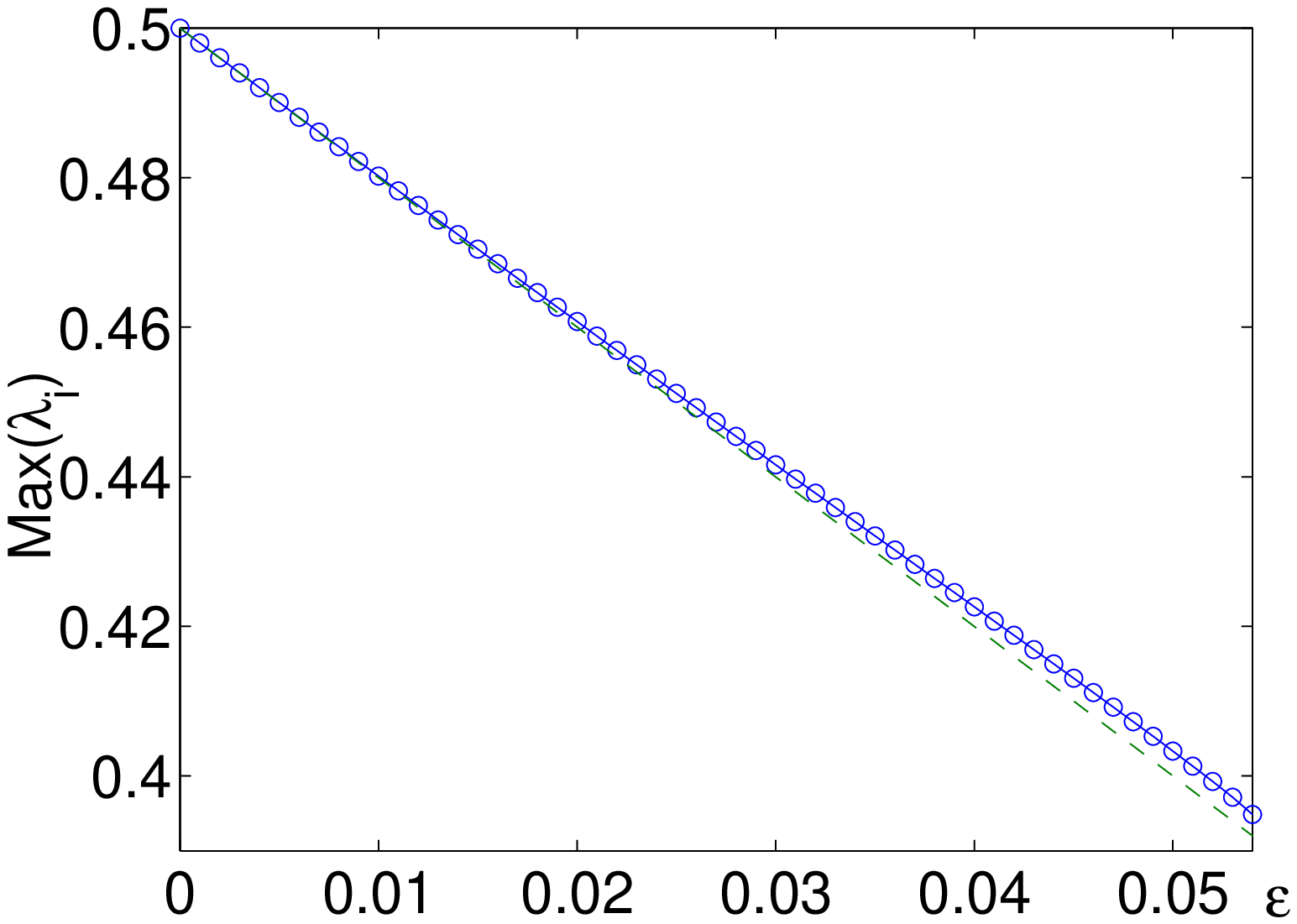}
\includegraphics[height=6cm,width=6cm]{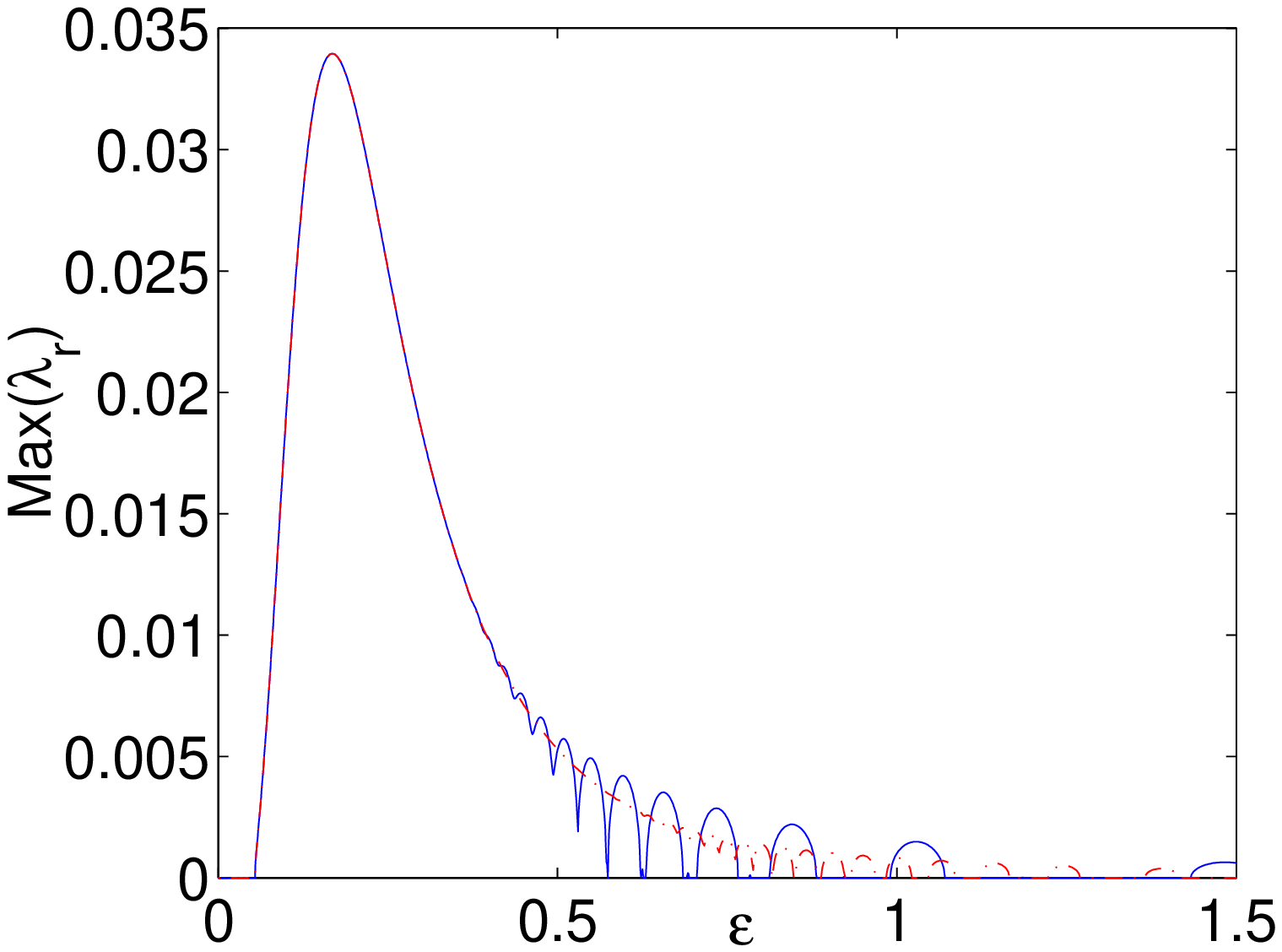} &  \\
\includegraphics[height=6cm,width=6cm]{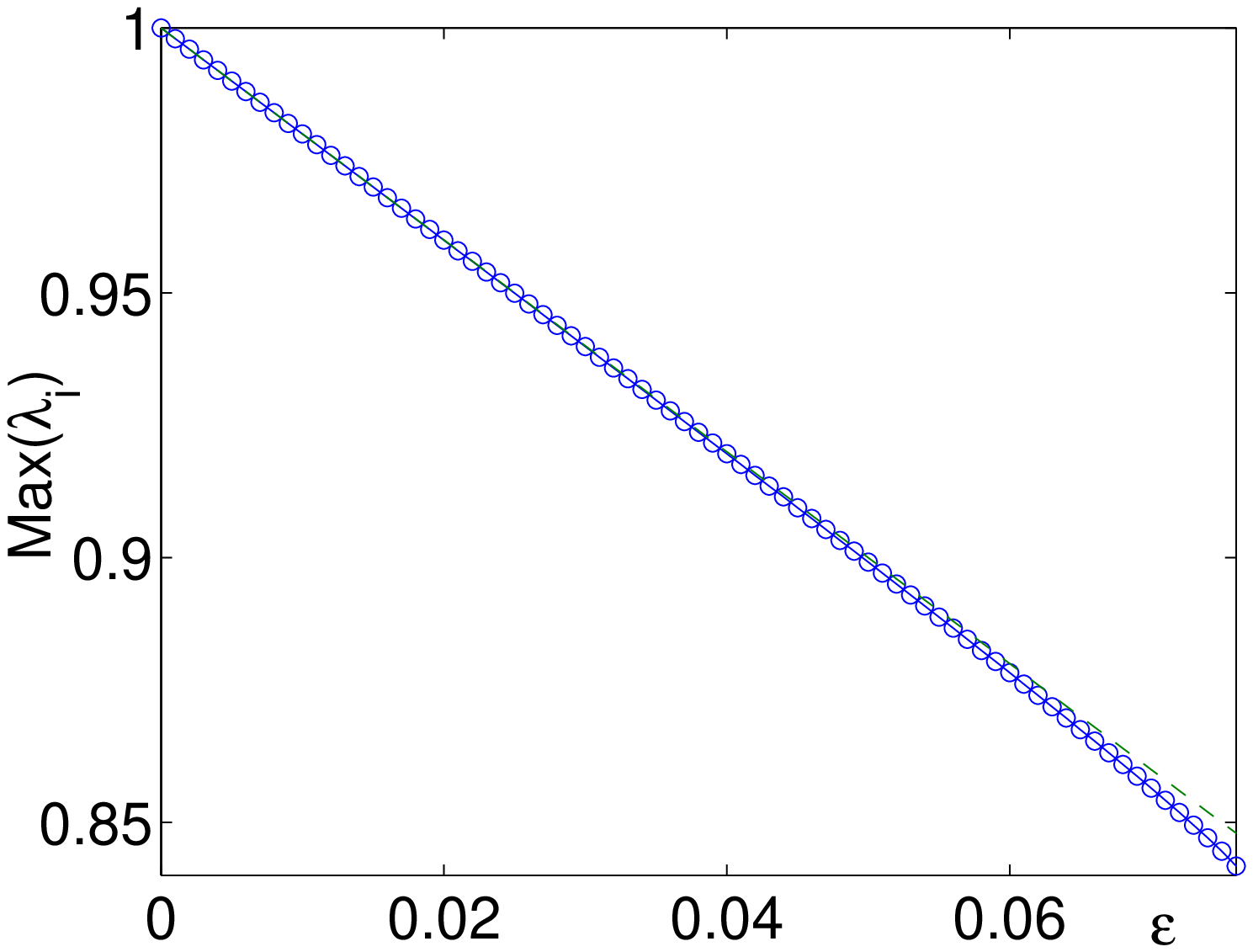}
\includegraphics[height=6cm,width=6cm]{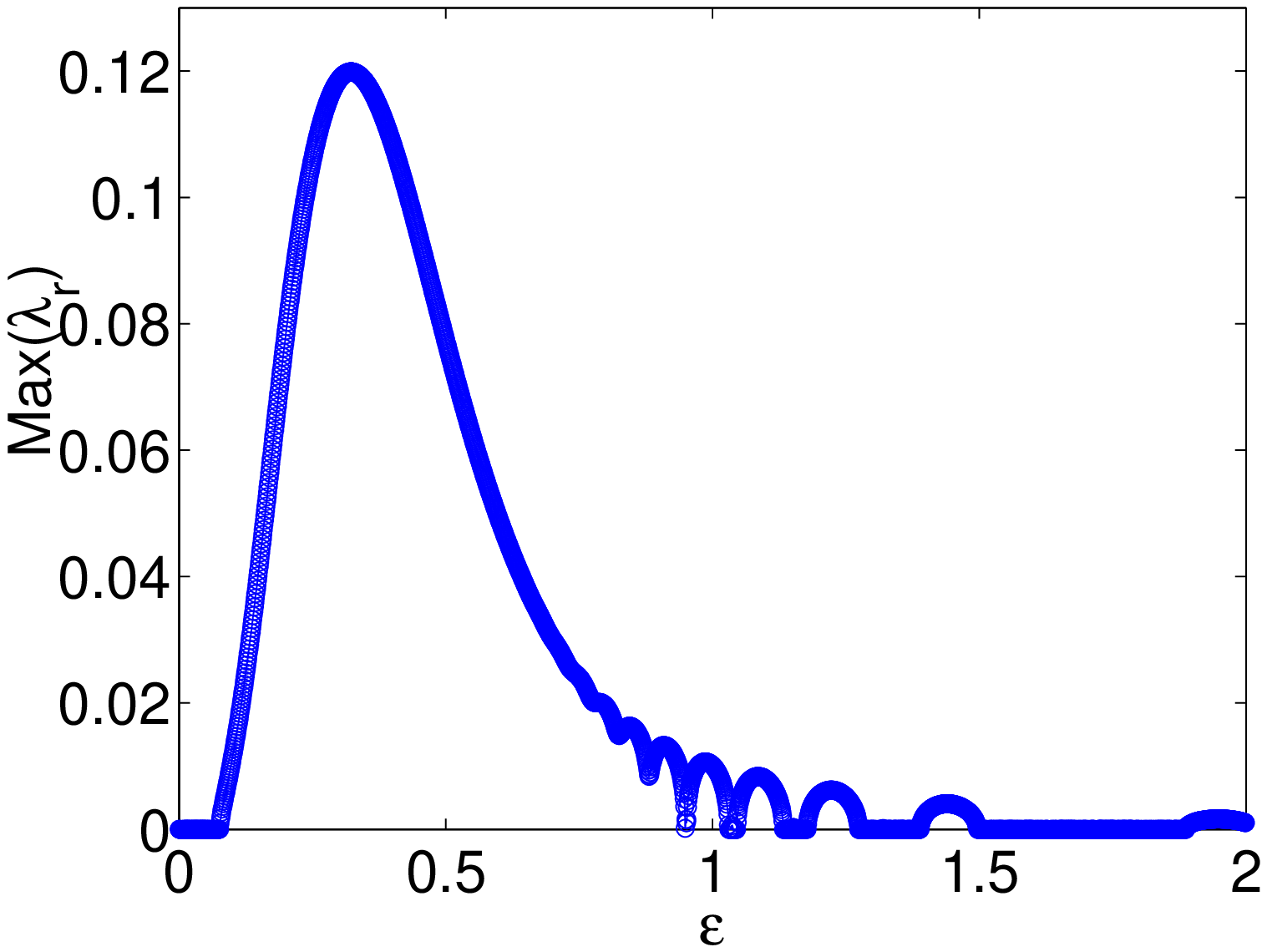} &  \\
\end{tabular}\end{center}
\par
\vskip-0.7cm 
\caption{(Color online) The figure shows stability results for the on-site dark soliton. 
The top panels pertain to the saturable model, while the bottom ones
to the cubic model. The left panels show the trajectory of the imaginary eigenvalue as a function
of $\epsilon$, starting at $i (\beta_2-\Lambda_2)$ and at $i \Lambda_1$ in the two models (the
maximal imaginary eigenvalue) until the point of collision $\epsilon_{cr}$
with the upper band edge of the continuous spectrum. The dashed lines
in both cases show the theoretical predictions of Eqs. (\ref{ceq8o}) 
and (\ref{ceq18}) respectively. The right panels show the real part
of the relevant eigenvalue, which is zero before --and non-zero
after-- the relevant collision due to the ensuing Hamiltonian Hopf
bifurcation leading to the exit of an eigenvalue quartet. In the
top (saturable) case, notice in addition to the solid line (computation
on a lattice with 250 sites) the dashed line which denotes the
same computation on a lattice with 1000 sites.}
\label{fig2}
\end{figure}

\begin{figure}[tbp]
\begin{center}
\hskip-0.15cm
\begin{tabular}{cc}
\includegraphics[height=6cm,width=6cm]{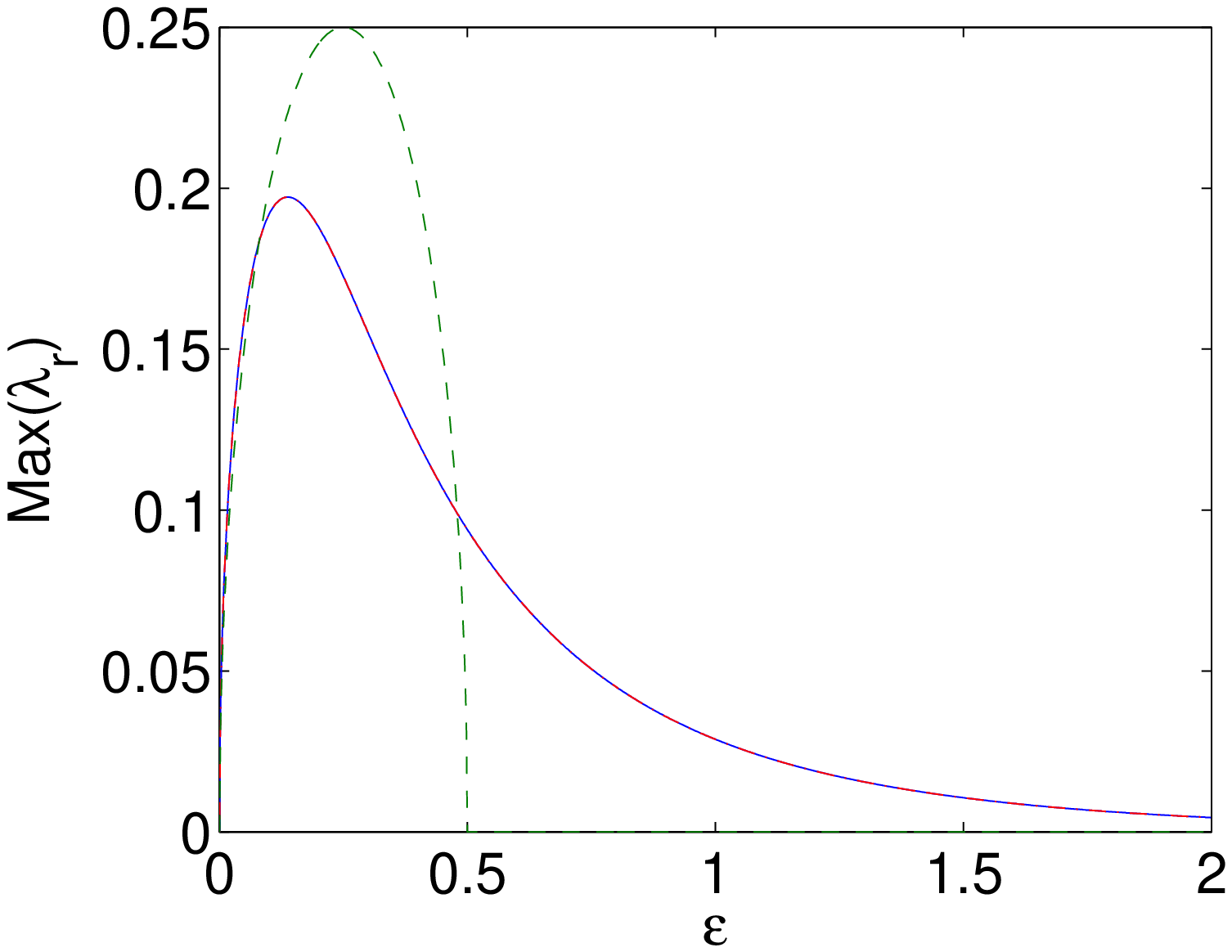}
\includegraphics[height=6cm,width=6cm]{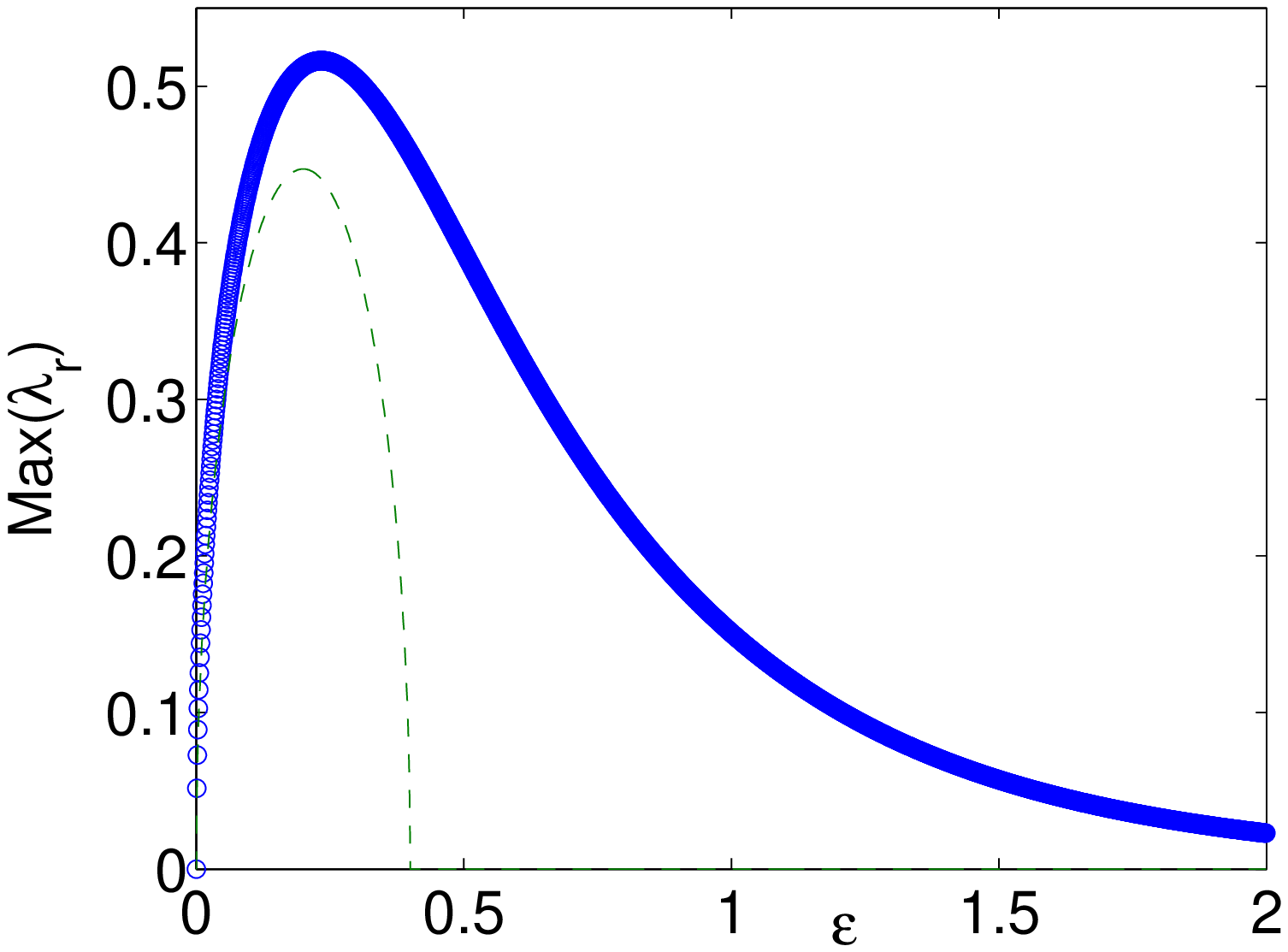} &  \\
\end{tabular}\end{center}
\par
\vskip-0.7cm 
\caption{(Color online) The figure shows the stability results for the inter-site dark soliton. 
The left and right panels correspond to the saturable and cubic nonlinearity cases, respectively. 
The solid lines refer to the numerical results, 
while the dashed ones to the theoretical predictions of Eqs. (\ref{ceq8m})
and (\ref{ceq16}), respectively. Notice that the configurations are
unstable $\forall \epsilon$ {\it in both cases}, 
due to the relevant real eigenvalue pair.}
\label{fig3}
\end{figure}

We now turn to a quantitative comparison of the stability results
for the on-site configurations in Fig. \ref{fig2} and the inter-site
configurations in Fig. \ref{fig3}. In the case of the on-site configurations,
we can clearly observe the eigenvalue starting from $i (\beta_2-\Lambda_2)$
in the saturable case and from $i \Lambda_1$ in the cubic one; 
this eigenvalue moves  
linearly (in a decreasing way) along the imaginary axis as $\epsilon$
is increased, as is predicted both by the Gerschgorin estimate, and
also remarkably accurately (see the dashed lines in the left panels
of Fig. \ref{fig2}) by the anti-symmetric approximation of
Eqs. (\ref{ceq8o}) and (\ref{ceq18}) respectively. This eigenvalue collides, in our
numerical results, with
the band edge of the continuous spectrum for
$\epsilon \approx 0.055$ in the saturable model and $\epsilon \approx 0.077$
in the cubic one; this is in remarkable agreement with the theoretical 
predictions of $0.0538$ for the saturable case and $0.07735$ for the cubic one,
respectively given by Eqs. (\ref{ceq8p}) and (\ref{ceq19}). Notice that
the latter case was the one examined previously in \cite{joh}.
After the collision, the Hamiltonian Hopf bifurcation ensues, resulting
into the emergence of a complex quartet of eigenvalues in the spectral
plane. This eigenvalue approaches the spectral plane origin at $\lambda=0$,
as the continuum limit of $\epsilon \rightarrow \infty$ is approached. 
Here, it is worth briefly iterating on one of the important points of 
\cite{joh} for reasons of completeness. Our computations, performed with 
lattices of $N=250$ lattice sites, seem to illustrate the presence of 
restabilization windows where the eigenvalue ``sneaks into'' the imaginary
axis as $\epsilon$ grows. However, this is a consequence of the 
finite computational size of the lattice, which leads to quantization of
the relevant wavenumbers $k$ and, as a result, to the presence of gaps
in the continuous spectral band of Eqs.  (\ref{ceq8k}) and (\ref{ceq14}).
This is also illustrated in the top right panel of Fig. \ref{fig2} featuring
(by a dashed line) a computation with a lattice four times larger than
the original one, where many of the original gaps have been eliminated.
Notice, however, that this issue (of finite size stabilization) may be 
one that is relevant to experimental situations as, e.g., in the experiments of \cite{kip} 
where propagation over 250 channels was reported (and, in fact, the generated beam was only 
about 25 channels wide). 

For the inter-site dark solitons, we show the relevant stability results 
in Fig. \ref{fig3}, both for the saturable (left panel) and the cubic (right panel) nonlinearity.
The theoretical predictions of Eqs. (\ref{ceq8m}) 
and (\ref{ceq16}) are also shown by dashed lines and provide a
fair approximation of the relevant eigenvalue, especially for
small $\epsilon$ (the perturbative result is not expected 
to be accurate for $\epsilon > 0.2$). An important observation
here is the fact that the real eigenvalue pair, predicted to 
bifurcate immediately for $\epsilon >0 $ for the inter-site configuration, 
remains {\it always} real up to the continuum limit, asymptotically 
approaching the spectral plane origin as $\epsilon \rightarrow \infty$. 
Interestingly, this is so {\it both} for the saturable and for
the cubic nonlinearity cases and this feature is unaffected by the lattice size
(we repeated the computation of the saturable case with a four
times larger lattice, only to find an identical pair of 
real eigenvalues). 

\begin{figure}[tbp]
\begin{center}
\hskip-0.15cm
\begin{tabular}{cc}
\includegraphics[height=6cm,width=6cm]{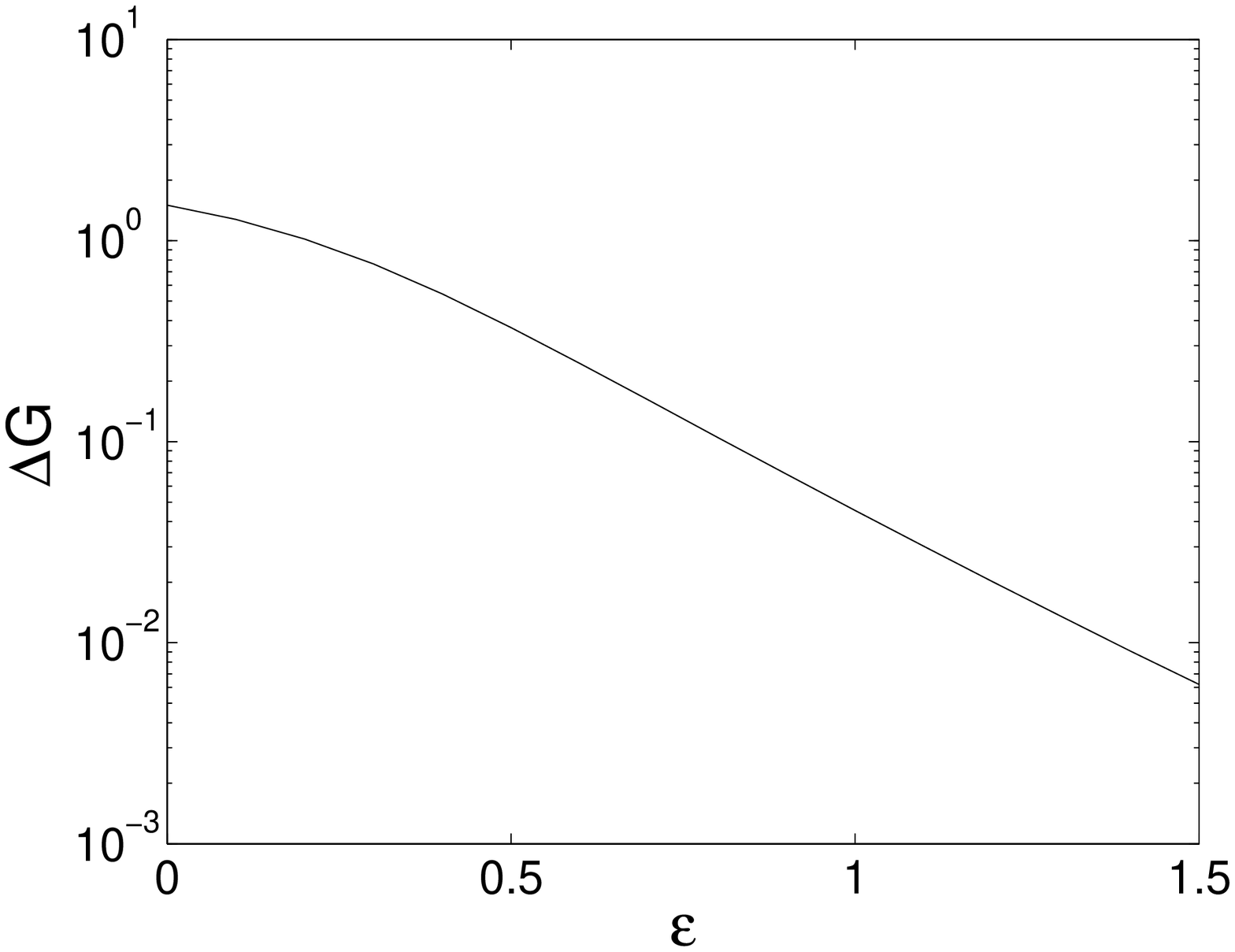}
\includegraphics[height=6cm,width=6cm]{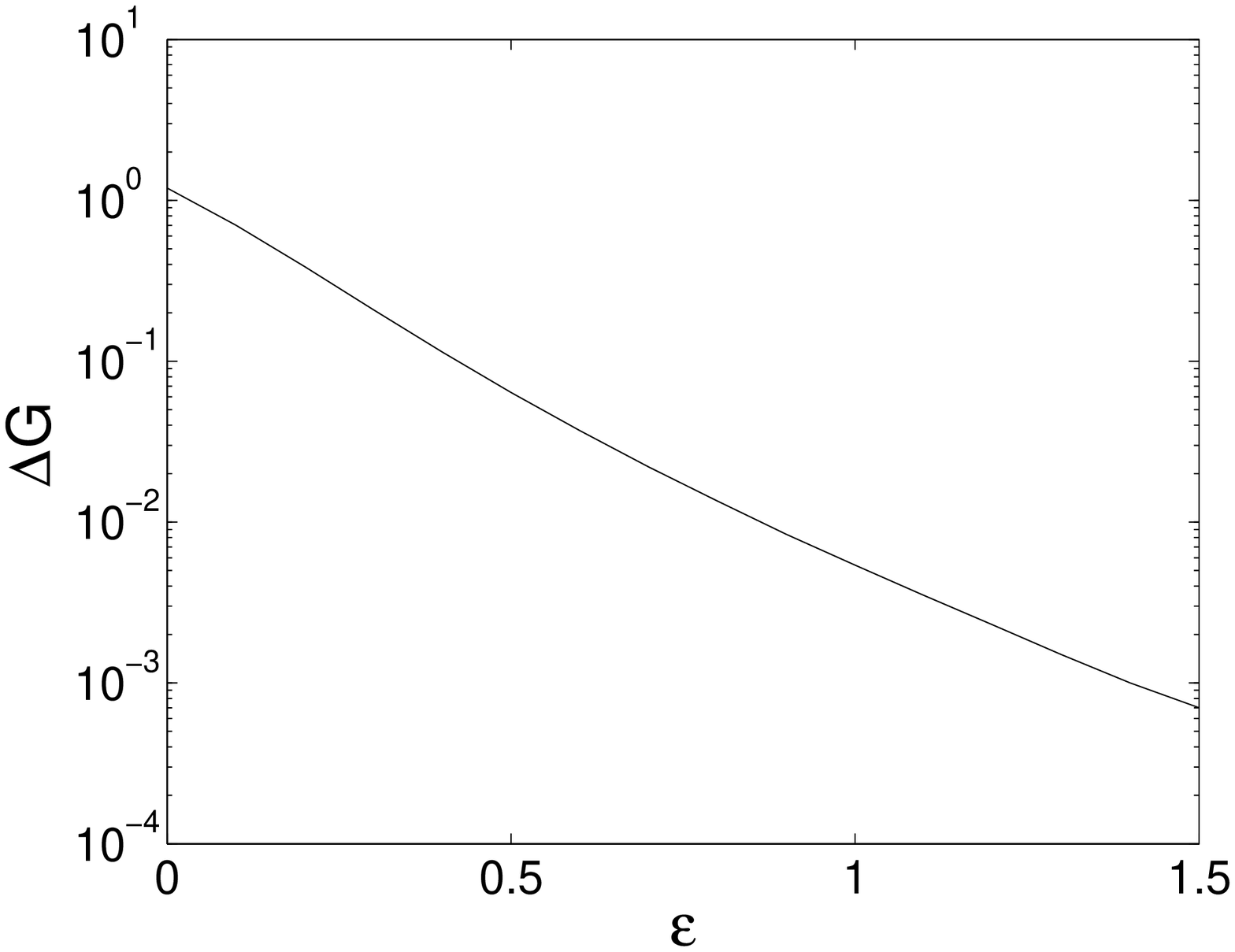} &  \\
\end{tabular}\end{center}
\par
\vskip-0.7cm 
\caption{The figure shows a semi-log plot of the difference $\Delta G \equiv G^{B}-G^{A}$ 
of the grand canonical energies $G^A$ and $G^B$ [see Eq. (\ref{G})] 
of the on-site and inter-site dark solitons, respectively, as a function of $\epsilon$. The left and right 
panels correspond to the cases of the cubic and the saturable nonlinearity, respectively.
}
\label{figG}
\end{figure}

The above result also highlights a noteworthy difference between 
the saturable model in its defocusing and focusing \cite{kip2} forms. 
The focusing version of the model has the remarkable feature of
stability alternation 
between on-site and inter-site configurations (vanishing of the so-called Peierls-Nabarro barrier) 
as one of its most important characteristics. This is apparently occurring \cite{hadz1} through
an exchange of a real eigenvalue pair (pertaining to one of the 
configurations) and of an imaginary pair (pertaining to the other) 
at ``exceptional'' values of $\epsilon$, marking these so-called 
transparency windows. Interestingly, this feature appears to be 
absent from the defocusing version of the model, possibly because 
of the structurally different nature of the spectrum and the 
presence of the Hamiltonian Hopf bifurcation, which does not allow
such an exchange of eigenvalue pairs to occur at any finite value
of $\epsilon$. We corroborate this point by considering  
the grand canonical free energy defined as 
\begin{equation}
G_j = H_j - \Lambda_{j}N, 
\label{G}
\end{equation}
where $j=1,2$ for the cubic and saturable nonlinearity cases, respectively, 
$N=\sum_{n} |u_{n}|^{2}$ is the power (or the number of atoms in the BEC context) and $H_j$ denotes 
the Hamiltonian of the respective models, which is given by 
\begin{equation}
H_1=\sum_{n} \left[ (u_{n+1}-u_{n})^{2} + \frac{ \beta_{1}}{2} |u_{n}|^4 \right],
\label{Hc}
\end{equation}
for the cubic nonlinearity, and  
\begin{equation}
H_2=\sum_{n} \left[ (u_{n+1}-u_{n})^{2} - \beta_{2} \ln(1+|u_{n}|^{2}) \right],
\label{Hp}
\end{equation}
for the saturable nonlinearity. $G$ (and, in particular, its differences
between on-site and inter-site configurations) was shown in \cite{hadz2}
to be the proper indicator of the location of the exchange of
stability/transparency windows in the focusing version of the model.
Therefore, in Fig. \ref{figG} we show 
the difference $\Delta G \equiv G^{B}-G^{A}$ of the grand canonical free
energies $G^A$ and $G^B$
of the on-site and inter-site dark solitons, as a function of $\epsilon$
in the two models. This quantity has 
the sense of a ``pinning energy barrier'' and is a measure of the 
well-known Peirls-Nabarro barrier. As is clearly seen in Fig. \ref{figG}, 
$\Delta G$ is a monotonically decreasing function of $\epsilon$ in 
{\it both} the cubic and 
saturable nonlinearity cases 
(shown, respectively, in the left and right panels of Fig. \ref{figG}); 
note that a similar result has been obtained in \cite{oksana} for 
the cubic case. 
This result is a clear indication that, in contrast to the saturable 
self-focusing nonlinearity case, in the self-defocusing regime the 
saturable lattice possesses {\it no transparency points}.
It remains as an interesting open question whether 
such a stability alternation could be achievable in other 
defocusing lattice models (such as, e.g., possibly a cubic-quintic 
equation), especially because these transparency windows have been 
associated with important travelling solution properties of these 
lattices \cite{hadz2}.

\begin{figure}[tbp]
\begin{center}
\hskip-0.15cm
\begin{tabular}{cc}
\includegraphics[height=6cm,width=6cm]{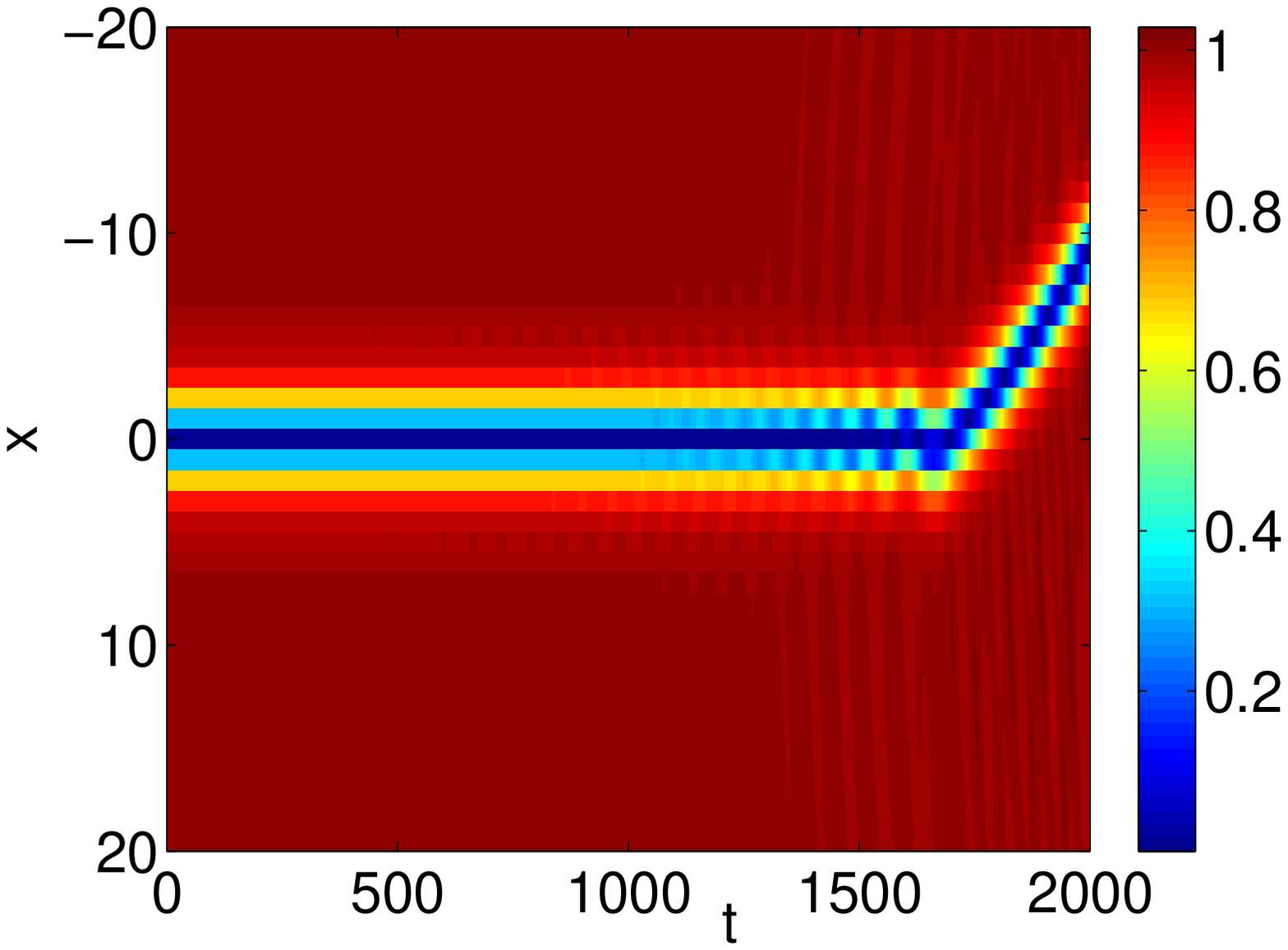}
\includegraphics[height=6cm,width=6cm]{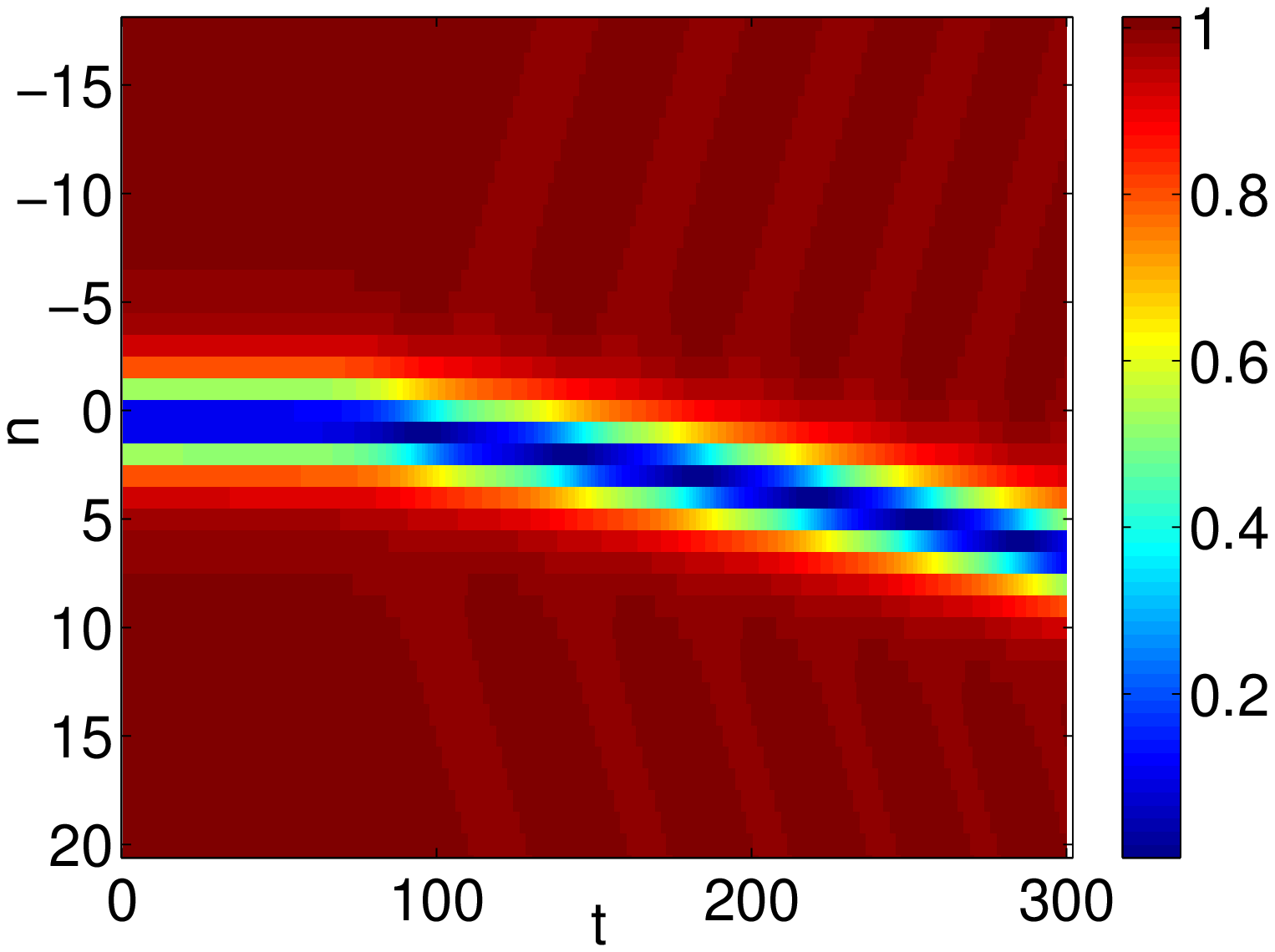} &  \\
\includegraphics[height=6cm,width=6cm]{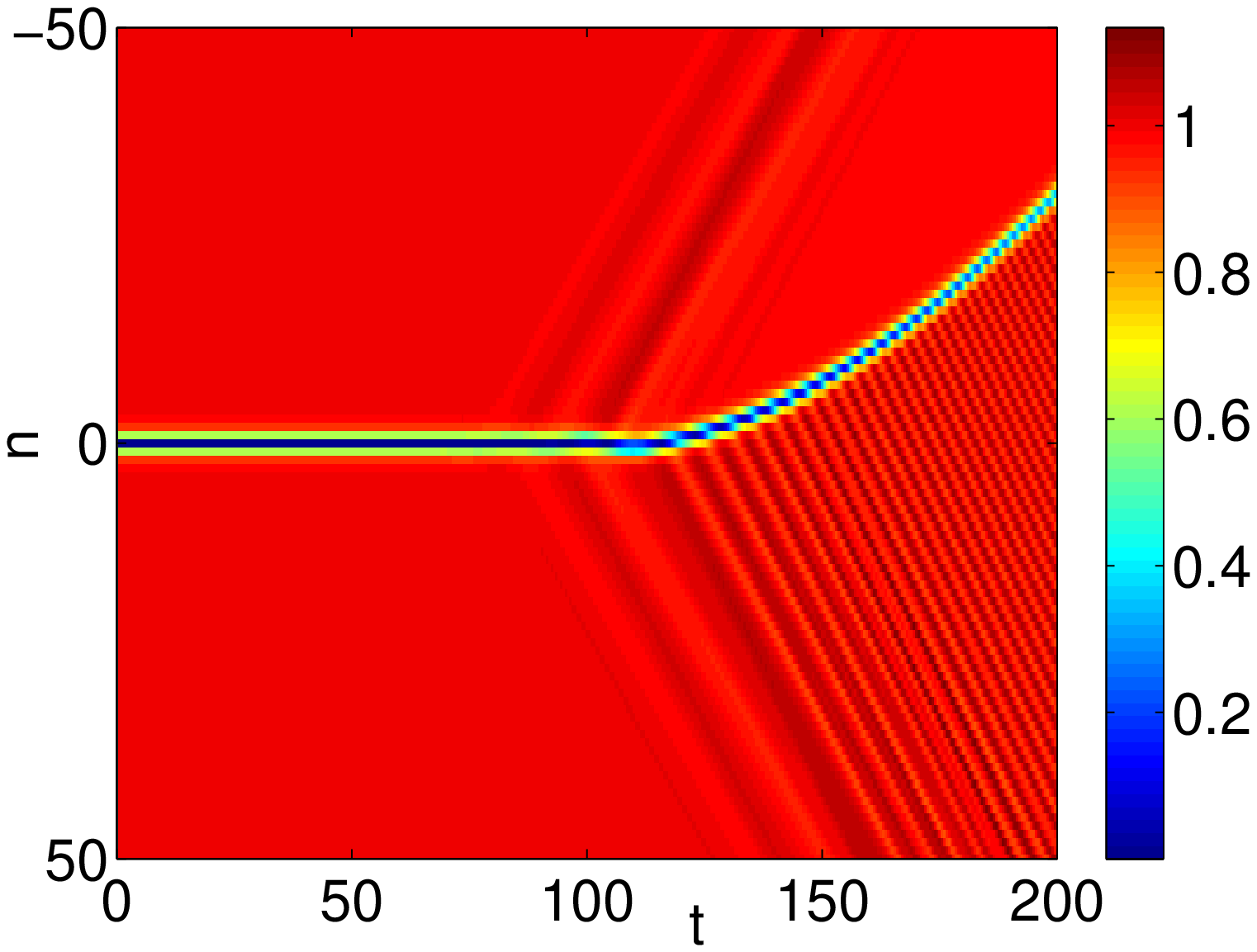}
\includegraphics[height=6cm,width=6cm]{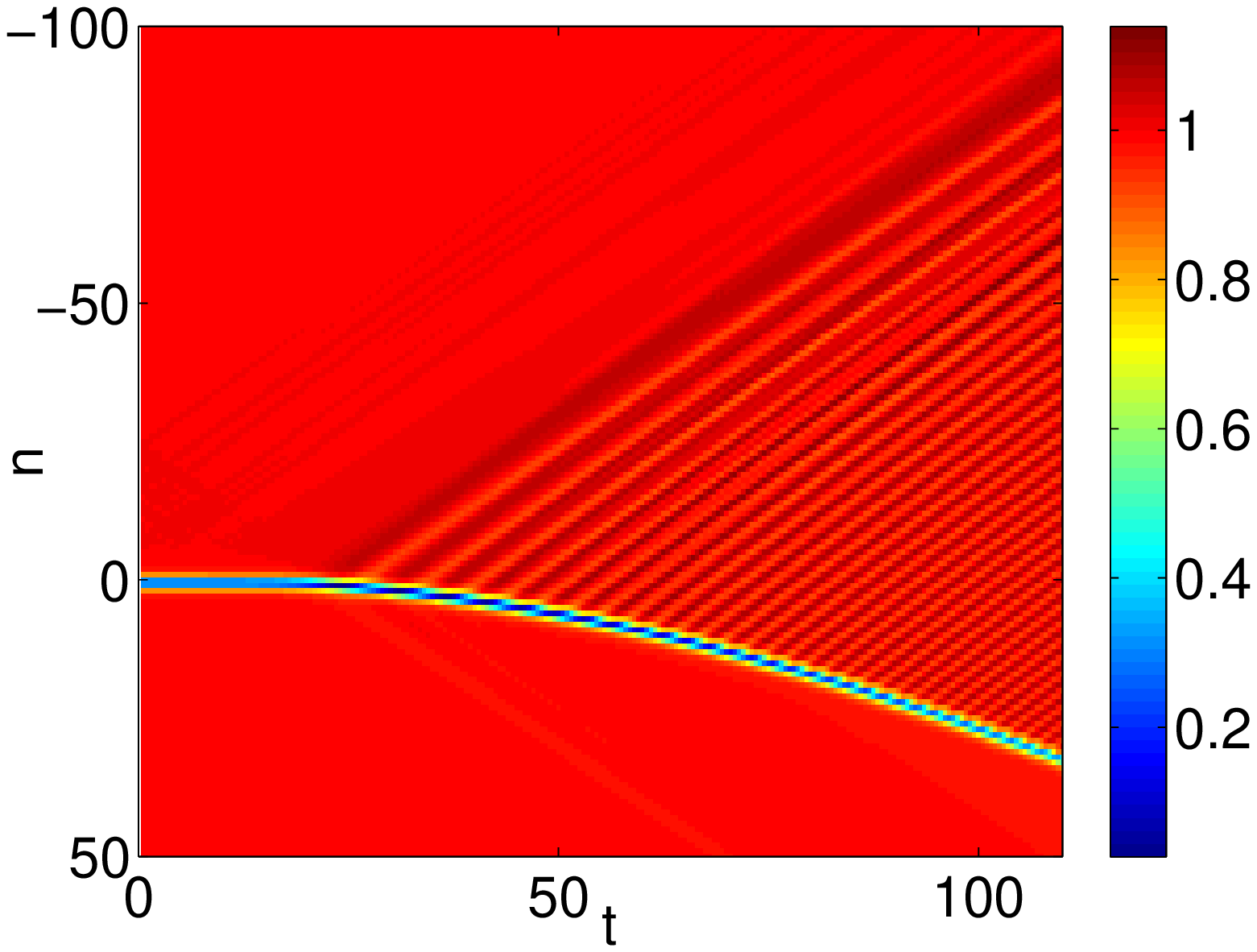} &  \\
\end{tabular}\end{center}
\par
\vskip-0.7cm 
\caption{(Color online) The evolution of the instability for the saturable
on-site (top left), saturable inter-site (top right), cubic on-site
(bottom left) and cubic inter-site (bottom right) dark solitons.
The dynamics is shown by a space-time contour plot of the square modulus
$|u_n(t)|^2$ of the field.}
\label{fig4}
\end{figure}

We also briefly compare the growth rates of the instabilities
in the different models. In the saturable case and for the on-site
configuration, the maximal growth rate (i.e., the maximal 
real part of the most unstable eigenvalue) is $\approx 0.0339$ and 
occurs for $\epsilon \approx 0.168$. On the other hand,
for the cubic case it is $\approx 0.1199$ (for $\epsilon \approx 0.322$).
For the inter-site case, the maximal rate is $0.1972$ for the saturable
case (and occurs for $\epsilon \approx 0.138$), while it is $0.5163$
for the cubic one (for $\epsilon \approx 0.234$). This illustrates
a general feature that we have observed (and which is justified on the
basis of the nature of the nonlinearity), namely that the more unstable
situations emerge for the ``stronger'' cubic nonlinearity in comparison
with its saturable counterpart. This is also evinced by the dynamical
evolution of the instability for a given $\epsilon$($=0.5$ in the case
of Fig. \ref{fig4}). In particular, the figure shows the spatio-temporal
contour plot evolution of the square modulus $|u_n|^2(t)$ of the field.
It is clear that the dark soliton of the saturable model (top panels) with the weaker
growth rates becomes unstable for longer times than its cubic counterpart.
Moreover, the inter-site dark soliton (with the real eigenvalue and the --generally--
larger growth rates than the eigenvalue quartet of the on-site case) 
of the right panels becomes unstable more rapidly than the on-site soliton 
of the left panels. The instability growth rates for this particular
value of $\epsilon$ are $0.0051$ for the on-site saturable case 
(top left), $0.0939$ for the inter-site saturable (top right),
$0.0774$ for the on-site cubic (bottom left) and $0.3944$ for the
inter-site cubic (bottom right). The ordering of the growth rates
can be readily (and inversely) associated with the time it takes
for the corresponding structures to become unstable. Finally, we
note in passing that the oscillatory instability is manifested
through an oscillation of the center, as is expected based on 
the relevant complex eigenvalue and its associated oscillatory 
growth. At the same time, the real eigenvalue pairs lead to a uni-directional
propagation of the dark soliton as a result of the instability.
This corroborates the results of \cite{oksana} for the cubic case.

\section{Conclusions and Future Challenges}

In the present paper, motivated by recent experiments
in waveguide arrays featuring a 
saturable self-defocusing nonlinearity \cite{kip},
as well as earlier experimental works \cite{silb} in 
arrays exhibiting cubic (Kerr) nonlinearities, we have examined
and compared two correspondingly relevant nonlinear dynamical
lattice models, namely the one with a saturable nonlinearity, as well
as the intensively examined one with the cubic nonlinearity.
The objective of the study was to obtain an analytical,
to the extent possible, understanding of the features of dark solitons, 
reported in earlier works, either computational or experimental ones
(such as those cited above).

Earlier reported features included, among others, the 
stable nature of on-site dark solitons versus the 
unstable nature of inter-site ones.
We rationalized this on the basis of an immediately unstable
(off of the anti-continuum limit) real eigenvalue pair in the
latter case, versus an eigenvalue on the imaginary axis, per
the zero site, in the former case. However, it was illustrated
(corroborating and expanding the earlier results of \cite{joh,oksana}),
how this imaginary eigenvalue leads to a Hamiltonian-Hopf bifurcation
eventually unveiling an oscillatorily unstable behavior in the dynamics
of the corresponding on-site dark solitons.

These earlier, including very recent, 
results also indicated (see e.g. \cite{kip})
an apparent stabilization of even inter-site modes in the saturable case (in comparison with the cubic one). 
We illustrated that while this does not appear to be due to a true stabilization of
the solutions, it is justified by the much weaker instability growth
rates of the saturable model, compared to its more strongly nonlinear cubic sibling. 

In addition to the above, we highlighted the usefulness of the 
anti-continuum limit and of perturbative expansions off of it,
in order to obtain not only qualitative but also quantitative
insight regarding the relevant structures and their form, as well as, more
importantly, their stability. 

These techniques can be readily applied to other defocusing
lattices as well and it would be of interest to do so. This is especially so in light
of the fact that the present models did not feature an apparent 
exchange of stability such as the one showcased by the focusing analog
of the saturable model. It would be interesting to examine whether
other models, such as ones with competing nonlinearities in the form
of, e.g., the cubic-quintic model might present such stability
alternations (possibly using the extra ``knob'' of manipulating
the strength of the quintic term). Of equal interest would be
the expansion of considerations developed herein into higher-dimensional
defocusing lattices and to a potential identification of the natural
multi-dimensional generalization of the dark soliton solutions presented
herein in the form of a vortex type structure. Such works are currently
in progress and will be reported in future publications.

\vspace{5mm}

{\bf Acknowledgements.} PGK gratefully acknowledges support from
NSF-DMS-0204585, NSF-DMS-0505663 and NSF-CAREER.

\appendix

\section{Perturbation theory using formal expansion}

In this appendix, as advertised above, 
we present another method of deriving the approximation 
to the key eigenvalues of discrete dark solitons both in cubic and 
saturable nonlinearities and for both inter-site and
on-site cases. Here we will use the mathematically familiar approach
of formal perturbation expansions.

\subsection{Cubic nonlinearity}

First, let us consider the simpler case, i.e., the cubic nonlinearity case.

To perform the linear stability analysis around the discrete solitary waves, 
we introduce the following linearization
ansatz
\[
u_n=v_n+ \delta C_n.
\]
Substituting into \eqref{ceq1}  yields the following linearized equation
to $\mathcal{O}(\delta)$
\begin{equation}
i\dot C_n=-\epsilon\Delta_2C_n+ \beta_1\left(2|v_n|^2C_n+v_n^2 C_n^\star\right)-\Lambda_1 C_n.
\label{lin}
\end{equation}

Decomposing $C_n(t)=\eta_n+i\xi_n$ and assuming that $v_n$ is real, 
eq.\ \eqref{lin} gives (see, e.g., \cite{joh})
\begin{eqnarray}
\left(\begin{array}{cc}
\dot{\eta}_n\\
\dot{\xi}_n
\end{array}\right)=
\left( 
\begin{array}{cc} 
0 & {\cal L}_-(\epsilon) \\
-{\cal L}_+(\epsilon) & 0
\end{array} 
\right)
\left(\begin{array}{cc}
{\eta}_n\\
{\xi}_n
\end{array}\right)=
{\cal H}
\left(\begin{array}{cc}
{\eta}_n\\
{\xi}_n
\end{array}\right),
\label{lin2}
\end{eqnarray}
where the operators ${\cal L}_-(\epsilon)$ and ${\cal L}_+(\epsilon)$ are defined as $\mathcal{L}_-(\epsilon)\equiv -\epsilon\Delta_2+ \beta_1v_n^2-\Lambda_1$ and $\mathcal{L}_+(\epsilon)\equiv -\epsilon\Delta_2+3 \beta_1v_n^2-\Lambda_1$. The stability of $v_n$ is then determined by the eigenvalues of ${\cal H}$.

Let the eigenvalues of ${\cal H}$ be denoted by $\lambda$, which implies that 
$v_n$ is stable if $\lambda_r=0$. Because \eqref{lin2} is linear, 
we can eliminate one of the eigenvectors, for instance
$\xi_n$, from which we obtain the following eigenvalue problem
\begin{equation}
{\cal L}_-(\epsilon){\cal L}_+(\epsilon)\eta_n=-\lambda^2\eta_n=\Xi\eta_n.
\label{evp}
\end{equation}

As before, we expand the eigenvector $\eta_n$ and the eigenvalue $\Xi$ as 
\[\eta_n=\eta_n^{(0)}+\epsilon\eta_n^{(1)}+{\cal O}(\epsilon^2),\quad \Xi=\Xi^{(0)}+\epsilon\Xi^{(1)}+{\cal O}(\epsilon^2).\]

Substituting into Eq.\ \eqref{evp} and identifying coefficients for consecutive powers of the small parameter $\epsilon$ 
yields
\begin{eqnarray}
\displaystyle \left[{\cal L}_-(0){\cal L}_+(0)-\Xi^{(0)}\right]\eta_n^{(0)}&=&0,\label{o0}\\
\displaystyle \left[{\cal L}_-(0){\cal L}_+(0)-\Xi^{(0)}\right]\eta_n^{(1)}&=&f,\label{o1}
\end{eqnarray}
with
\begin{equation}
\displaystyle f=\left[(\Delta_2-2 \beta_1v^{(0)}_nv^{(1)}_n){\cal L}_+(0)+{\cal L}_-(0)(\Delta_2-6v_n^{(0)}v_n^{(1)})+\Xi^{(1)}\right]\eta_n^{(0)}.
\label{f}
\end{equation}

First, let us consider the order ${\cal O}(1)$ equation \eqref{o0}.
One can do a simple analysis as above to show that there is 
one eigenvalue, i.e., 
$\Xi^{(0)}=0$ for the inter-site configuration and two eigenvalues $\Xi^{(0)}=0$ and $\Xi^{(0)}=\Lambda_1^2$ for the on-site one.
The zero eigenvalue has infinite multiplicity and is related to 
the continuous spectrum, as discussed previously. 

For the inter-site configuration, there is an eigenvalue bifurcating from the 
continuous spectrum as soon as the coupling is turned on. Therefore, this 
zero eigenvalue is the crucial eigenvalue for its stability. The normalized 
eigenvector of this eigenvalue is $\eta_n^{(0)}=1/\sqrt2$, for 
$n=0,\,1$ and $\eta_n^{(0)}=0$ otherwise. For the on-site configuration, 
the crucial eigenvalue for the stability is $\Xi^{(0)}=\Lambda_1^2$ with the 
normalized eigenvector $\eta_n^{(0)}=1$, for $n=0$ and 
$\eta_n^{(0)}=0$ otherwise.


The continuation of the critical eigenvalue when the coupling $\epsilon$ is turned on can be calculated
from Eq.\ \eqref{o1}. Due to the fact that the corresponding eigenvector is zero almost everywhere, we only need to 
consider the site with non-zero component eigenvector, i.e.\ $n=0,\,1$ for the inter-site and $n=0$ for the on-site. 

It is simple to show that the solvability condition of Eq.\ \eqref{o1} using, e.g., 
the Fredholm alternative requires $f=0$ from which 
one immeadiately obtains that $\Xi^{(1)}=-2\Lambda_1$ for the inter-site and $\Xi^{(1)}=-4\Lambda_1$ for the inter-site. 

Hence, the critical eigenvalue is:
\begin{equation}
\lambda=\pm \sqrt{2\Lambda_1\epsilon+{\mathcal O}(\epsilon^2)}
\label{result1}
\end{equation}
for the inter-site configuration and 
\begin{equation}
\lambda=\pm i \sqrt{\Lambda_1^2-4\Lambda\epsilon+{\mathcal O}(\epsilon^2)}
\label{result2}
\end{equation}
for the on-site one. 

These two results agree with the leading order calculations of
section II.A. In particular, Eq. (\ref{result1}) can be immediately
seen to agree with the leading order prediction of Eq. (\ref{ceq16}).
On the other hand, as regards Eq. (\ref{result2}), a Taylor 
expansion to leading order yields 
$\lambda = \pm i (\Lambda_1 -2 \epsilon)$, again  in agreement with the
findings of Eq. (\ref{ceq18}).

\subsection{Saturable nonlinearity}

For the saturable nonlinearity, substituting 
the same spectral ansatz $u_n=v_n+ \delta C_n$ 
into \eqref{ceq20} yields the following linearized equation
to $\mathcal{O}(\delta)$
\begin{equation}
i\dot C_n=-\epsilon\Delta_2C_n+ \frac{\beta_2}{1+|v_n|^2}\left( C_n-\frac{v_n(v_nC_n^\star+v_n^\star C_n)}{1+|v_n|^2} \right)-\Lambda_2 C_n.
\label{linn}
\end{equation}

With this nonlinearity, the operator ${\cal L}_-(\epsilon)$ and ${\cal L}_+(\epsilon)$ are now defined as 
$\mathcal{L}_-(\epsilon)\equiv -\epsilon\Delta_2+\beta_2(1+v_n^2)/(1+|v_n|^2)^2-\Lambda_2$ and $\mathcal{L}_+(\epsilon)\equiv -\epsilon\Delta_2+\beta_2(1-v_n^2)/(1+|v_n|^2)^2-\Lambda_2$.

Again, by using 
$\Xi=\Xi^{(0)}+\epsilon\Xi^{(1)}$ and 
$\eta_n=\eta_n^{(0)}+\epsilon\eta_n^{(1)}$, we obtain the same equations as Eqs.\ (\ref{o0}-\ref{o1}) with the inhomogeneity term $f$ now given by
\begin{equation}
\displaystyle f=\left[(\Delta_2-k_{1n}){\cal L}_+(0)+{\cal L}_-(0)(\Delta_2-k_{2n})+\Xi^{(1)}\right]\eta_n^{(0)},
\label{f1}
\end{equation}
with 
\[
k_{1n}=-\frac{2\beta_2v_n^{(0)}v_n^{(1)}}{{(1+{v_n^{(0)}}^2)}^2},\quad k_{2n}=-\frac{2\beta_2v_n^{(0)}v_n^{(1)}}{{(1+{v_n^{(0)}}^2)}^2}
\left[
\frac{2(1-{v_n^{(0)}}^2)}{(1+{v_n^{(0)}}^2)}+1
\right].
\]

The stability of the static solutions of Eqs.\ \eqref{ceq5b}-\eqref{ceq7c} and \eqref{ceq8e}-\eqref{ceq8i} can then be determined using exactly the same procedures as above. One can simply calculate that the critical eigenvalue is
\begin{equation}
\lambda=\pm \sqrt{\frac{2\Lambda_2}{\beta_2}(\beta_2-\Lambda_2)\epsilon+{\mathcal O}(\epsilon^2)}
\label{result3}
\end{equation}
for the inter-site configuration and
\begin{equation}
\lambda=\pm i \sqrt{(\beta_2-\Lambda_2)\left((\beta_2-\Lambda_2)-4\epsilon\right)+{\mathcal O}(\epsilon^2)}
\label{result4}
\end{equation}
for the on-site configuration.

The eigenvalue prediction from Eq. (\ref{result3}) can be immediately 
seen to be identical to the leading order prediction of Eq. (\ref{ceq8m}).
On the other hand, Taylor expanding the result of Eq. (\ref{result4}),
we obtain $\lambda = \pm i (\beta_2-\Lambda_2 -2 \epsilon)$, which 
once again is identical to the prediction of Eq. (\ref{ceq8o}).

\end{document}